\documentclass[prc,aps,nofootinbib,%
tightenlines,superscriptaddress,floatfix,
preprintnumbers]{revtex4}
\usepackage{bbm}
\usepackage{graphicx}
\usepackage{amsmath}
\usepackage{amsfonts,amsbsy}
\usepackage{amssymb}
\usepackage{color}

\newcommand{\dsigmap}{{\frac{\ud \sigma^\textrm{p}_\textrm{dip}}{\ud^2 \bt}}}

\newcommand{\ud}{\, \mathrm{d}}

\newcommand{\rt}{{\mathbf{r}_\perp}}

\newcommand{\bt}{{\mathbf{b}_\perp}}
\newcommand{\st}{{\mathbf{s}_\perp}}
\newcommand{\kt}{{\mathbf{k}_\perp}}

\newcommand{\be}{\begin{equation}}
\newcommand{\ee}{\end{equation}}
\newcommand{\bea}{\begin{eqnarray}}
\newcommand{\eea}{\end{eqnarray}}

\begin{document}
\title{QCD saturation at the LHC: comparisons of models to p+p and A+A data and predictions for p+Pb collisions}
\author{Prithwish Tribedy}
\affiliation{Variable Energy Cyclotron Centre, 1/AF Bidhan Nagar, Kolkata-700064, India}
\author{Raju Venugopalan}
\affiliation{Physics Department, Bldg. 510A, Brookhaven National Laboratory, Upton, NY 11973, USA}

\begin{abstract}
In a previous paper (arXiv:1011.1895), we showed that saturation models, constrained by e+p HERA data on inclusive and diffractive cross-sections, are in good agreement with p+p data at LHC in the soft sector. Particularly impressive was the agreement of saturation models with the multiplicity distribution as a function of $n_{\rm ch.}$. In this paper, we extend these studies further and consider the agreement of these models with data on bulk distributions in A+A collisions. We compare our results to data on central and forward particle production in d+Au collisions at RHIC and make predictions for inclusive distributions in p+Pb collisions at the LHC. 

\end{abstract}
\preprint{ }
\maketitle
\section{Introduction}
In a previous paper~\cite{Tribedy:2010ab}, we discussed the computation of inclusive distributions in p+p collisions at RHIC and the LHC within the framework of $k_\perp$ factorization, wherein the unintegrated gluon distributions were determined from fits to small $x$ HERA data on inclusive, diffractive and exclusive final states. The key ingredient in these fits is the dipole cross-section, 
which, to leading logarithmic accuracy, can be defined as
\begin{equation}
\dsigmap(\rt,x,\bt)=
2\,\left(1-\frac{1}{N_c}\,\Big< {\rm tr}\, \big({\tilde U}(\bt + \frac{\rt}{2}){\tilde U}^\dagger(\bt-\frac{\rt}{2})\big)\Big>_x\;\right)\; ,
\label{eq:Wilson-amp}
\end{equation}
where ${\tilde U}(\bt \pm \frac{\rt}{2})$ is a Wilson line in the fundamental representation representing the interaction between a quark and the color fields of the target. In the Color Glass Condensate (CGC) framework~\cite{MV,CGC-reviews} of gluon saturation~\cite{GLR}, the average $\langle\cdots\rangle_x$ is an average over these color fields; the energy dependence of the correlator as a function of $x$ (or the rapidity $Y=\ln(1/x)$) is given by the JIMWLK equation~\cite{JIMWLK}.  In the large $N_c$ limit, the equation for the energy evolution of this correlator is the Balitsky-Kovchegov (BK) equation~\cite{BK}.

We note however that neither JIMWLK nor BK is at present equipped to deal well with the impact parameter dependence of the dipole 
cross-section; the dipole cross-section in this formalism is taken in eq.~(\ref{eq:Wilson-amp}) to be independent of the impact parameter. Another limitation of this framework is that the full next-to-leading logarithmic (NLL) expressions are not yet available; at the NLL level, only 
running coupling corrections to the leading log kernel have been considered in phenomenological applications. With these limitations in mind, we considered in ref.~\cite{Tribedy:2010ab}, saturation models of the dipole cross-section with the common criteria that their parameters be strongly constrained by fits to the HERA data. The saturation models considered\footnote{For more details, see the discussion in section II of ref.~\cite{Tribedy:2010ab}.} included the IP-Sat model~\cite{KT}, the b-CGC model~\cite{IIM,KMW,KW} and the rcBK model~\cite{Albacete}. All of these models provide good fits to the HERA data\footnote{The rcBK model has only been compared to inclusive HERA data~\cite{arXiv:1012.4408}. We note however that this comparison is to the combined H1-ZEUS inclusive data~\cite{arXiv:0911.0884}, in contrast to the IP-Sat and b-CGC models, which were fit only to the older ZEUS data~\cite{ZEUS}. }.

In hadron-hadron collisions, one can derive at leading order the expression~\cite{BGV1}
\begin{equation}
{
\frac{\textmd{d}N_{g}(\textbf{b}_{\bot})}{\textmd{d}y~\textmd{d}^{2}\textbf{p}_{\bot}}=\frac{ 4 \alpha_S}{\pi C_F} \frac{1}{p_{\bot}^2} 
\int \frac{\textmd{d}^{2} \textbf{k}_{\bot}}{(2\pi)^{5}} \int \textmd{d}^{2} \textbf{s}_{\bot} \frac{\textmd{d}\phi_A(x_1,\textbf{k}_{\bot}|\textbf{s}_{\bot})}{\textmd{d}^2\textbf{s}_{\bot}} \frac{\textmd{d}\phi_B(x_2,\textbf{p}_{\bot}-\textbf{k}_{\bot}|\textbf{s}_{\bot}-\textbf{b}_{\bot})}{\textmd{d}^2\textbf{s}_{\bot}} \, .
}
\label{eq:ktfact1}
\end{equation}
This equation is a generalization of the well known $k_\perp$ factorization expression for inclusive gluon production~\cite{Braun} to include the impact parameter dependence of the unintegrated gluon distributions. {Here $C_F = (N_c^2-1)/2 N_c$ is the Casimir for the fundamental representation.} Using a relation between quark and gluon dipole amplitudes strictly valid in the  large $N_c$ limit, the unintegrated gluon distribution in either of the two protons can be expressed in terms of the corresponding dipole cross-section measured in DIS as~\cite{GelisSV} 
\begin{equation}
\frac{\textmd{d}\phi(x,\textbf{k}_{\bot}|\textbf{s}_{\bot})}{\textmd{d}^2\textbf{s}_{\bot}} =\frac{\textbf{k}_\bot^2 N_c}{4 \alpha_S}  \int \limits_{0}^{+\infty}\textmd{d}^2\textbf{r}_{\bot}
e^{i \textbf k_{\bot}.\textbf{r}_{\bot}} \left[1 - \frac{1}{2}\, \frac{\ud \sigma^\textrm{p}_\textrm{dip}}{\ud^2 \textbf{s}_\perp\\} (\rt,x,\textbf{s}_\perp)\right]^{2}
\label{eq:unint-gluon}
\end{equation}
Thus the impact parameter dependent dipole cross-section determined from HERA data can be used to compute single inclusive gluon distributions in proton-proton collisions. Since this is a leading order computation, the overall normalization is not constrained and is determined from data as we shall describe below. For the integrated multiplicities, there is a logarithmic infrared divergence that can be regulated by introducing a mass term as discussed in our previous paper. We should mention that solutions of Yang-Mills equations that treat the infrared behavior properly give infrared finite distributions~\cite{hep-ph/9909203,hep-ph/0305112,{hep-ph/0303076}}. For a nice comparison of theoretical errors in various $k_\perp$ factorized approximations to the full classical Yang-Mills results, see ref.~\cite{arXiv:1005.0955}.

In ref.~\cite{Tribedy:2010ab}, we used eq.~(\ref{eq:ktfact1}) combined with eqs.~(\ref{eq:unint-gluon}) and (\ref{eq:Wilson-amp}) to compute the 
rapidity and $p_\perp$ distributions for the models discussed for central rapidities in p+p collisions at RHIC energies all the way to the highest 
available LHC energies. A typical feature of the rapidity and $p_\perp$ distributions in these models was that the agreement with data improved with increasing energy; this should be the case because saturation effects are increasingly important at higher energies. We also observed that the saturation models, in particular the IP-Sat model, gave excellent fits to the multiplicity distribution $P(n_{\rm ch.})$ as a function of  $n_{\rm ch.}$ for a wide range of energies. In this work, we will first briefly consider p+p collisions again before discussing A+A collisions and p/d+A collisions. In the latter case, we will present predictions for a future p+Pb run at the LHC. 

\section{Results for p+p collisions}
\begin{figure}[h]
\includegraphics[width =7cm, height =6.5cm]{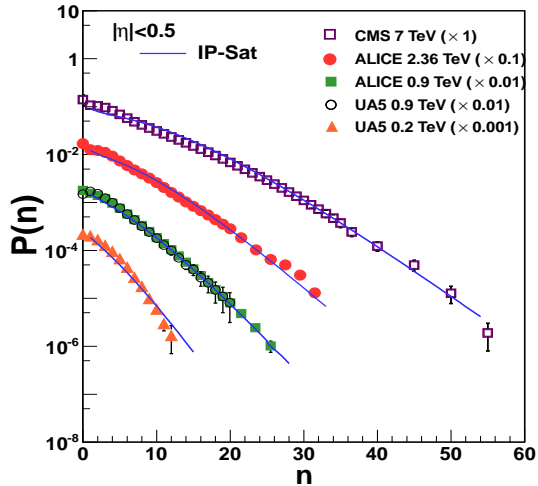}
\caption{Multiplicity distribution for p+p collisions in the IP-Sat model compared to  data from UA5, CMS and ALICE~\cite{ua53,Khachatryan:2010nk, alice2}.}
\label{fig:multdist_pp}
\end{figure}

The probability distribution for producing $n$ particles is
\begin{equation}
P(n) = \int d^2 \bt {dP_{\rm inel.} \over d^2\bt}\, P_n^{\rm NB}({\bar n}(\bt), k(\bt))\,,
\label{eq:mult-dist.}
\end{equation}
where, in the Glasma flux tube framework~\cite{DumitruGMV}, we can derive the negative binomial distribution~\cite{GelisLM}
\begin{equation}
P_n^{\rm NB}({\bar n}, k) = \frac{\Gamma(k+n)}{\Gamma(k)\Gamma(n+1)}\,\frac{{\bar n}^n k^k}{({\bar n} +k)^{n+k}} \, .
\label{eq:NB}
\end{equation}
with the parameter $k$ defined specifically to be 
\begin{equation}
k(\bt) = \zeta {(N_c^2-1) Q_S^2 S_\perp \over 2\pi} \, ,
\label{eq:flux1}
\end{equation}
with $ \zeta$=0.155 obtained from a fit to p+p multiplicity distribution. Here 
\begin{equation}
Q_S^2\, S_\perp (\bt) = \int d^2 s_\perp Q_S^2 (s_\perp, \bt ) \, , \nonumber
\end{equation}
where, motivated by CGC computations on dilute-dense collisions~\cite{hep-ph/0105268}, we choose $Q_S$ in the overlap area of the two hadrons to be $Q_S(s_\perp, \bt) = {\rm min.}\,\{Q_S(s_\perp), Q_S(s_\perp-\bt)\}$. Also, ${\bar n}$ is the 
average multiplicity at a given impact parameter in the saturation model. Finally, as previously, we use
\begin{equation}
{dP_{\rm inel.}^{\rm dip.}\over d^2 \bt} = {\frac{dN_g}{dy} (\bt) \over \int d^2\bt \frac{dN_g}{dy}(\bt)} \, .
\label{eq:imp-prob-dip}
\end{equation}
In fig.~(\ref{fig:multdist_pp}), we show results for the p+p multiplicity distribution plotted for $\sqrt{s}$ up to 7 TeV for $|\eta|<$0.5. This extends the comparison in ref.~\cite{Tribedy:2010ab} to the CMS data at 7 TeV, with good agreement to $n_{\rm ch.}=60$, nearly twice the range considered previously. The parameter $ \zeta$=0.155 is extracted from a fit to data at 0.9 TeV and used for all other energies\footnote{To avoid confusion, the value for $\zeta$ quoted in ref.~\cite{Tribedy:2010ab} is for $Q_S$ in the fundamental representation; $k(\bt)$ is identical to that discussed here. Our estimations in the adjoint representation are shown in ref.~\cite{arXiv:1101.5922}.}.  It is the only free parameter in our fit; however, this quantity was recently computed non-perturbatively {\it ab initio} by solving Yang-Mills equations numerically~\cite{LappiSV} for two gluon correlations from gauge fields generated in the collision of two dense color sources. The results of the numerical computation vary depending on parameter choices in the range $\zeta\sim 0.3$-$1.5$--the lower end of this range is therefore a factor of two larger than the best fit value. Given the many uncertainties in the computation, the agreement is quite good, keeping in mind that nothing a priori prevents $\zeta$ from being orders of magnitude different. The good agreement of this framework with the LHC data over several decades, taken at face value,  leads us to conjecture\footnote{To confirm this conjecture, one would need to demonstrate that the results are valid for instance in the rcBK framework. However, because the impact parameter dependence of inclusive distributions in this model is rather simplistic and not 
constrained by data, this is difficult to confirm meaningfully at present.}  that fluctuations in the number of produced gluons for a fixed distribution of hot spots gives a larger contribution to the multiplicity distribution than fluctuations in the distribution of hot spots themselves. The latter is of course only treated at the mean field level here in contrast to ``pomeron loop" contributions~\cite{Kovner:2011pe} --suggesting perhaps that the latter are suppressed. Because Glasma flux tubes generate long range rapidity correlations,  and can explain the distribution of high multiplicity events, our result provides further evidence corroborating  computations in this framework~\cite{Dumitru-etal,Dusling:2012ig} that suggest Glasma flux tubes generate the near side ridge seen in high multiplicity events by the CMS collaboration~\cite{CMS-ridge}. 

\begin{figure}[h]
\includegraphics[width =7cm, height =6.5cm]{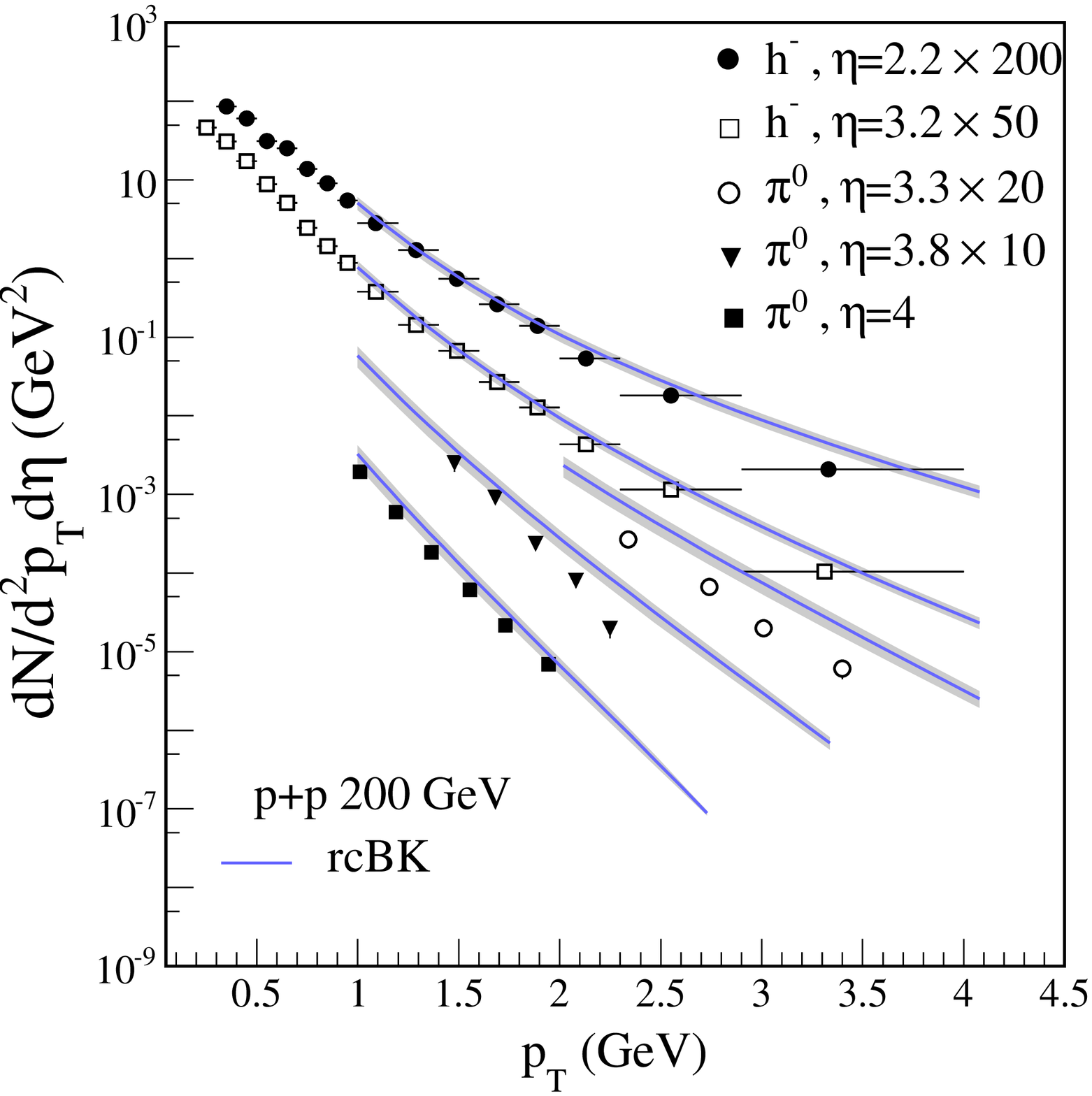}
\includegraphics[width =7cm, height =6.5cm]{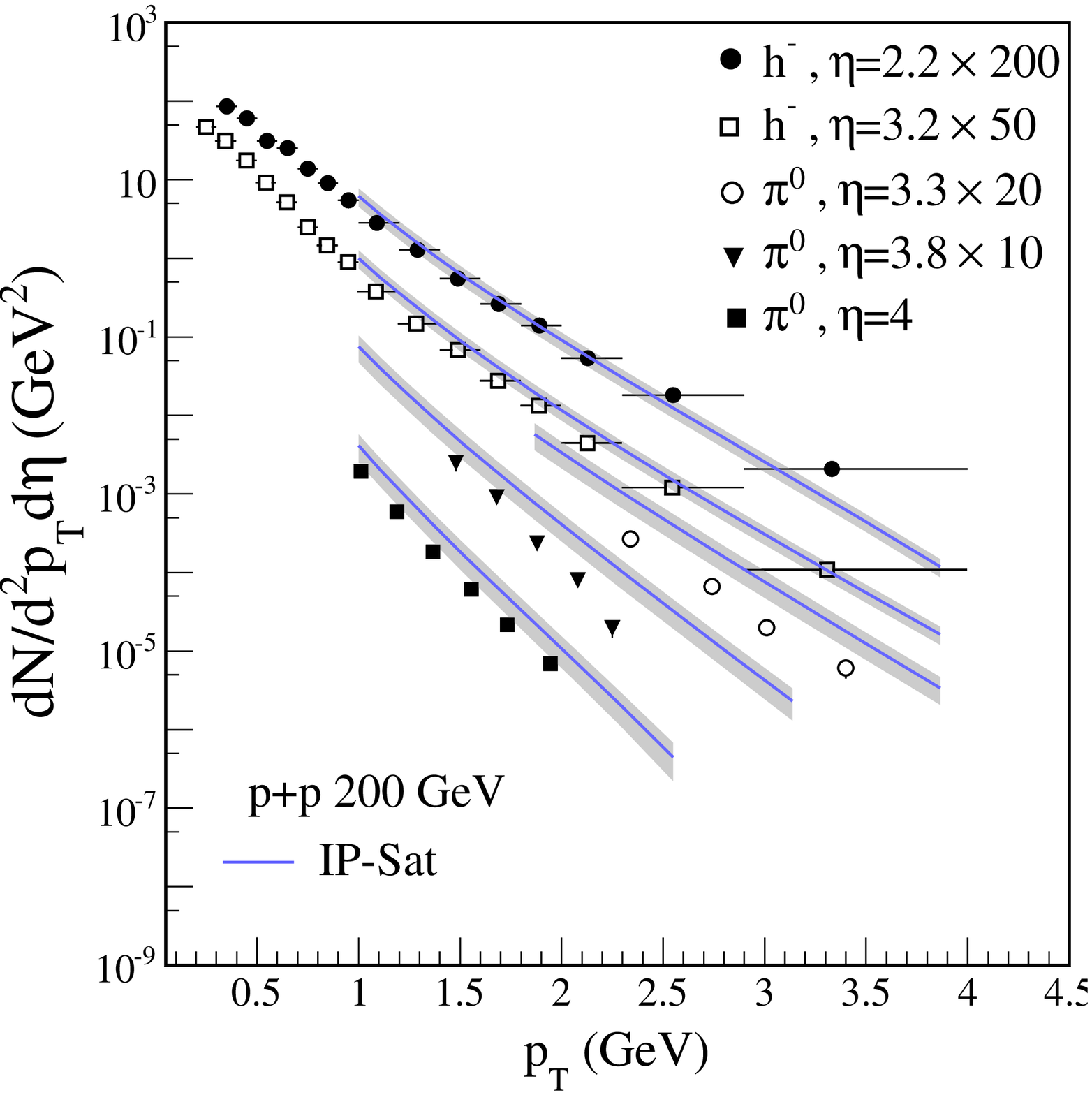}
\caption{Transverse momentum distributions at forward rapidities in rcBK and IP-Sat models compared to STAR~\cite{nucl-ex/0602011}  and BRAHMS~\cite{nucl-ex/0403005} data. The gray bands show the uncertainty in the determination of the normalization constant.}
\label{fig:ptdist_pp}
\end{figure}

We now turn to transverse momentum distributions for charged hadrons and $\pi^0$'s, that in p+p collisions are computed from the expression
\begin{equation}
\frac{d{\bar N}_{h}}{d^2p_{\bot}dy} = \int_{z_{\rm min.}}^1 \frac{dz}{z^2}\, \frac{d{\bar N}_{g}}{d^2q_{\bot}dy}\,D_{g\rightarrow h}\left(z=\frac{p_\perp}{q_\perp}, \mu^2\right) \,,
\label{eq:single-inclusive}
\end{equation}
where\footnote{The functional form quoted is for $\pi^+ + \pi^-$; we assume that the likelihood to fragment is identical for either charge and is equal to that for $\pi^0$. For other charged hadrons, the functional form is assumed to be identical with the normalization constrained by the momentum sum rule.} $D_{g\rightarrow h}(z,\mu^2)$ is chosen to be $6.05\,z^{-0.714}(1-z)^{2.92}$. In ref.~\cite{Tribedy:2010ab}, we had considered the transverse momentum distributions only at mid-rapidity, where agreement of the IP-Sat and rcBK models was not particularly good. This agreement improved significantly at the higher LHC energies. The explanation provided there was that this better agreement is a consequence of smaller $x$ values being probed at the higher energies. In fig.~(\ref{fig:ptdist_pp}), we show the results of the rc BK and IP-Sat models for forward rapidities for RHIC p+p collision at 200 GeV. 

{
For both IP-Sat and rcBK we have used an overall normalization\footnote{The overall normalization for minimum bias p+p is slightly different from our previous paper~\cite{Tribedy:2010ab} due to one less parameter ($\lambda_0$) in the large $x$ extrapolation which is set to zero here.} extracted from energy dependence of the single inclusive multiplicity of the form:
$A / (\pi  b_{\rm max}^2)$ with $b_{\rm max}= b_0+ C~ln(\sqrt{s})$. This form absorbs the uncertainties in the inelastic cross-section and higher order effects ($K$-factors). 
For the IP-Sat (rcBK) model, one finds $A=0.23 (6.15)$, $b_0= 5.77 (5.14)$ and $C= 0.32 (0.76)$ using the mass term $m=0.4$ GeV by fitting data points at $\eta=0$ over the range of energy shown in fig.~\ref{fig:avgmult_AA}. For the rc-BK model the constant term $A$ absorbs the prefactors 
\footnote{Eq.~\ref{eq:ktfact1} for min-bias p+p collision the rc-BK model normalization constant includes the constant prefactor ($\frac{4 \alpha_S}{\pi C_F (2\pi)^{5}} \frac{S_{A,B}}{(\pi R_A^2) (\pi R_B^2)}$);  here $R_A$, $R_B$ corresponds  to the radii of two protons and $S_{A,B}$ is the overlap area.} 
of Eq.~\ref{eq:ktfact1} which includes unknown overlap area and other terms that cannot be separated from the `` K factor ". } The results are rather insensitive to the infrared cut-off $m$. 
However, one finds a $\sim10\%$ variation of the normalization when the constants are extracted by a) fitting the full pseudo rapidity at RHIC energy, b) considering data points from different experiments. This variation, along with the numerical uncertainties, contributes to the gray bands shown in fig.~\ref{fig:ptdist_pp}.
We see that the agreement at forward rapidities is significantly better than our previous comparison to the mid-rapidity distribution; this result provides a good benchmark for computing $R_{pA}$ at RHIC and in predictions of the same for p+Pb collisions at the LHC. 

%

\section{Results for A+A collisions}

For a large nucleus, in the IP-Sat model, we can approximate the dipole-nucleus cross section to be
\begin{equation}
 \frac{\ud \sigma^\textrm{A}_\textrm{dip}}{\ud^2 \textbf{s}_\perp\\}  \approx 2\left[1-\exp\left\{-\frac{AT_A(\textbf{s}_\perp)}{2} \sigma_\textrm{dip}^p(\rt, x)\right\}\right]
\label{eq:nuc-dipole}
\end{equation}
where $\sigma_\textrm{dip}(\rt, x)^p$ is obtained from integrating the dipole-proton cross section in eq.~(\ref{eq:Wilson-amp}) over the impact parameter distribution in the proton. This form of the dipole-nucleus cross-section was shown previously~\cite{KLV} to give reasonable fits to the limited available fixed target e+A inclusive data. The initial conditions for rcBK evolution for a nucleus were similarly fixed by comparisons to the e+A data~\cite{arXiv:0911.2720}.
%

Substituting the expression for the dipole-nucleus cross-section in eqs.~(\ref{eq:ktfact1}), and likewise the latter in (\ref{eq:unint-gluon}), one can compute the nuclear multiplicity distributions. The infrared divergence in the multiplicity distribution is regulated in exactly the same was as was the case for the p+p multiplicity distribution, by replacing $p_{\perp }$ by $m_\perp=\sqrt{p_\perp^2+m^2}$, with $m$ varied between $0.2$-$0.4$ GeV. Wherever we have considered fixed coupling, we have used $\alpha_S$=0.2; for the running coupling case, we run $\alpha_S$ with the scale $Q_S = {\rm max.}\,\{Q_S(x_1, s_\perp), Q_S(x_2, s_\perp-\bt)\}$.

\begin{figure}[h]
\includegraphics[width =7cm, height =7cm]{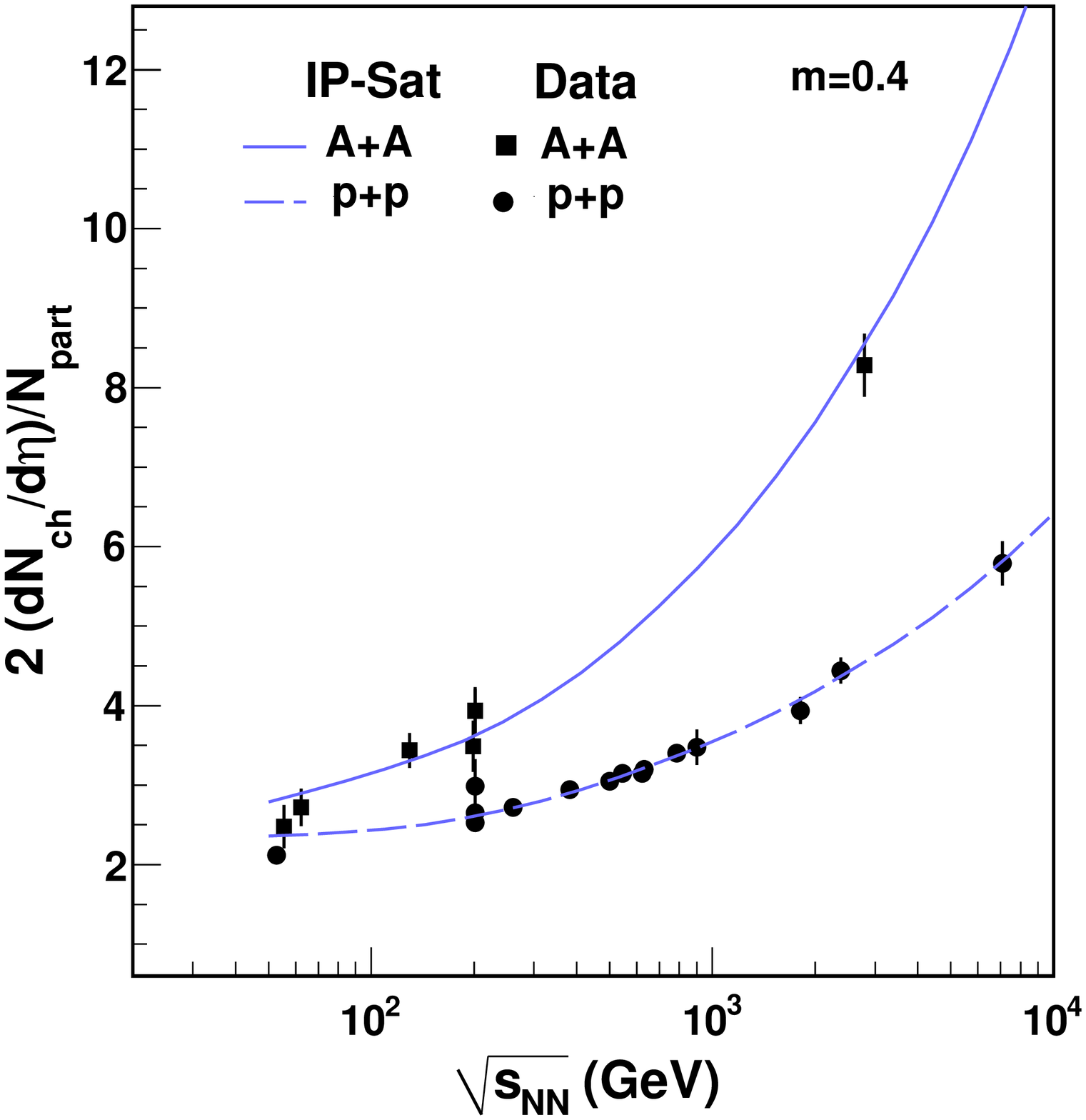}
\includegraphics[width =7cm, height =7cm]{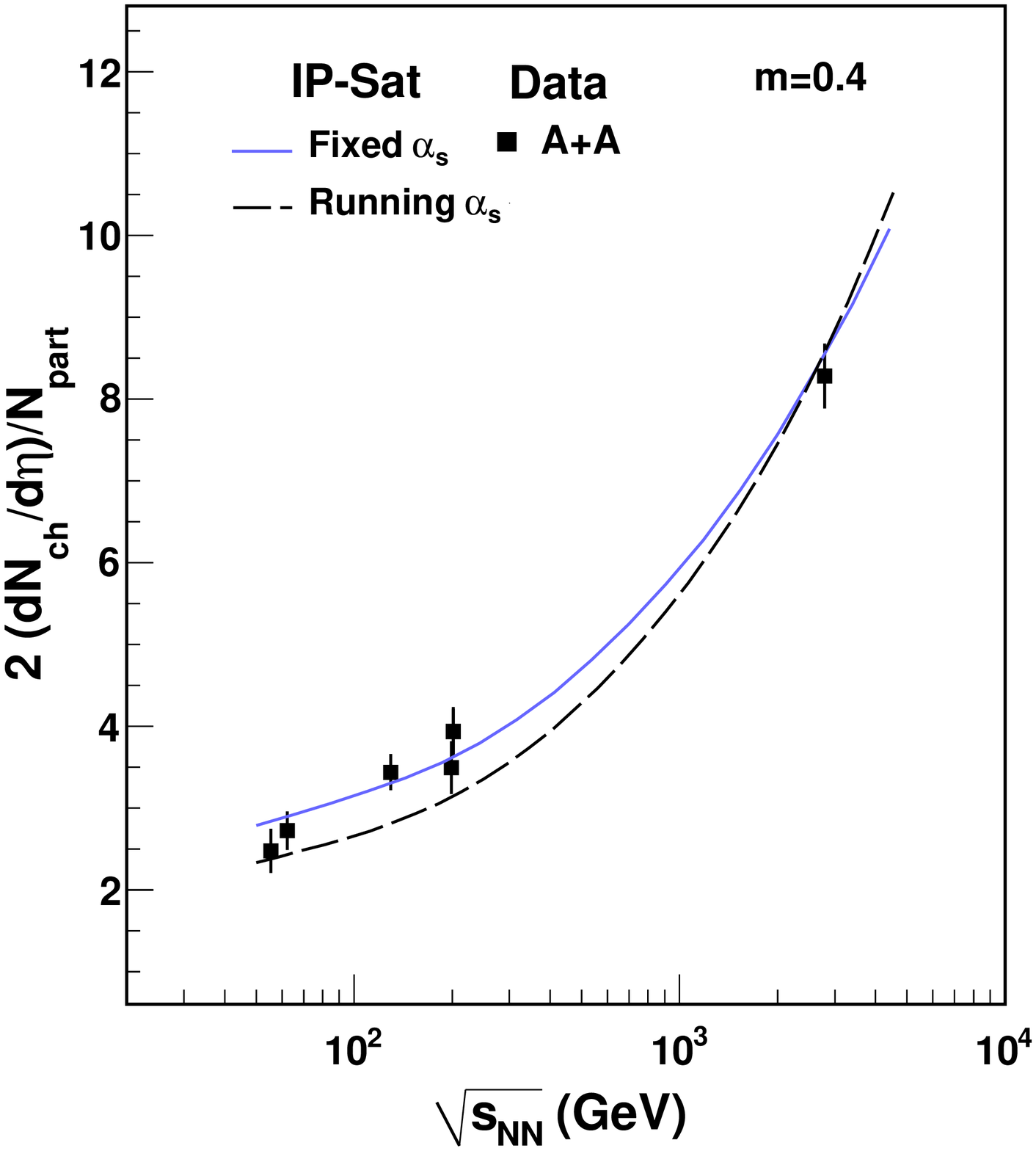}
\caption{Left: Energy dependance of the multiplicity per participant in the IP-Sat model for p+p and A+A collisions. For the A+A case, the calculation is done for the $0-6\%$ centrality. 
Right: same plot for A+A with fixed (solid) and running (dashed) coupling. Data points for p+p are from ref.~\cite{arXiv:1004.3034, arXiv:1002.0621, pp_mult_1, pp_mult_2} and for A+A from ref.~\cite{arXiv:0808.2041,AA_mult}}
\label{fig:avgmult_AA}
\end{figure}

Fig.~(\ref{fig:avgmult_AA}) shows the energy dependence of average multiplicity for most central Au+Au collision for fixed and running coupling in the IP-Sat model. 
The number of participants\footnote{In this expression, $\sigma_{\rm NN} \sim$ 62 mb for 2.76 TeV and $\sigma_{\rm NN} \sim$ 41 mb at 200 GeV.} at a given impact parameter is determined from the Glauber relation~\cite{Kharzeev-Nardi}
\be
N_{part}(\bt) \,=\, A \int T_A(\st) \left\{ 1-[1-T_B(\st \!-\bt) \sigma_{\rm NN} ]^B  \right\} d^2\st  \, +\,  B \int T_B(\st) \left\{ 1-[1-T_A(\st\!-\bt) \sigma_{\rm NN} ]^A  \right\} d^2\st \nonumber 
\ee
The results shown in fig.~(\ref{fig:avgmult_AA}) are for $0$-$6\%$ centrality, which corresponds to a median $\bt\approx 12.2$ GeV$^{-1}$; we compute ${dN_{\rm ch.}(\bt) \over d\eta}$ and 
$N_{\rm part}(\bt)$ for this median value. 
We observe that a fairly good agreement with data is obtained for the infrared cut-off given by $m=0.4$ GeV. The prescription for the running coupling gives a variation that corresponds to a $20\%$ uncertainty at lower energies, which decreases significantly at higher energies.

\begin{figure}[h]
\includegraphics[width =7cm, height =7cm]{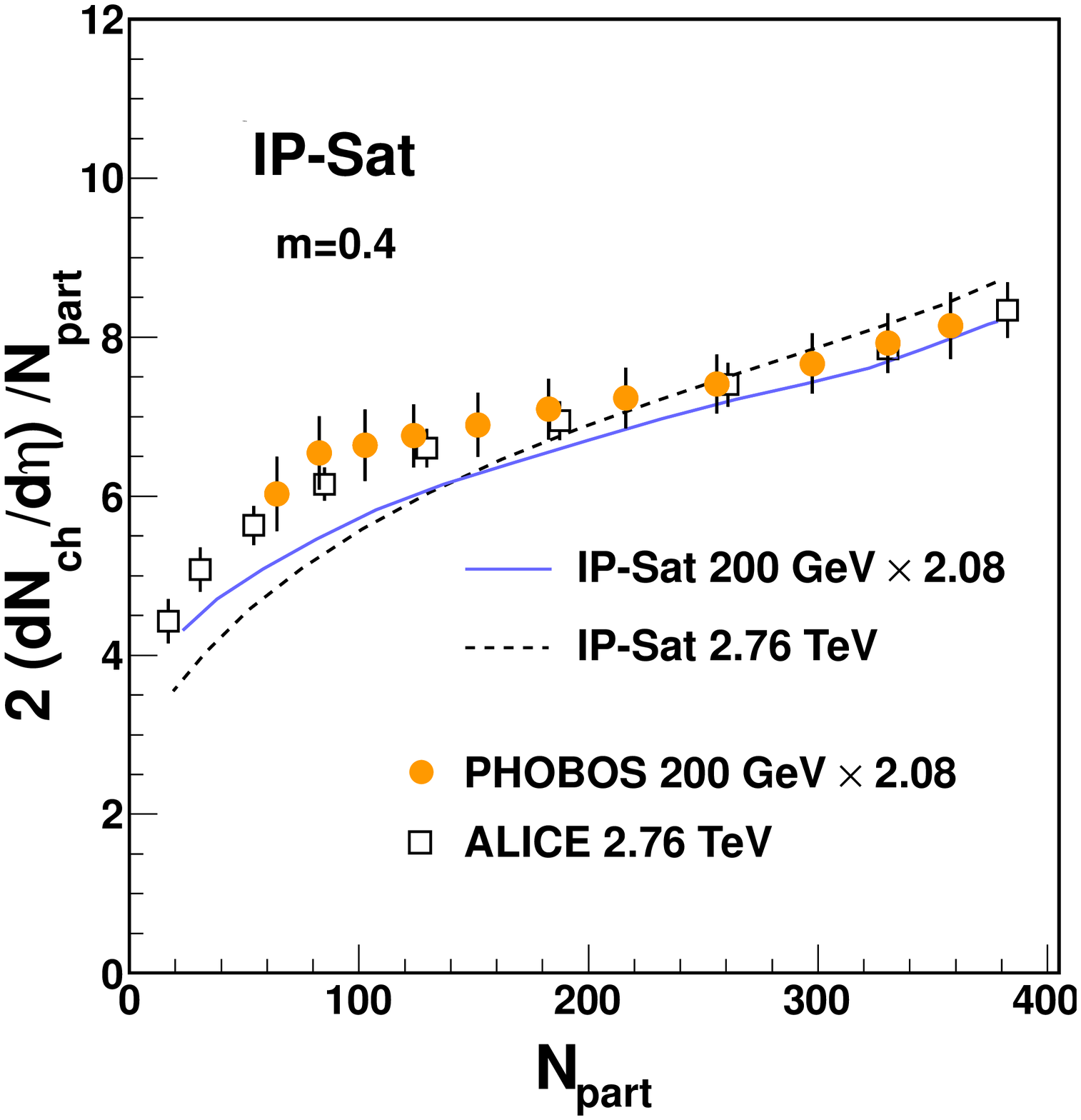}
\includegraphics[width =7cm, height =7cm]{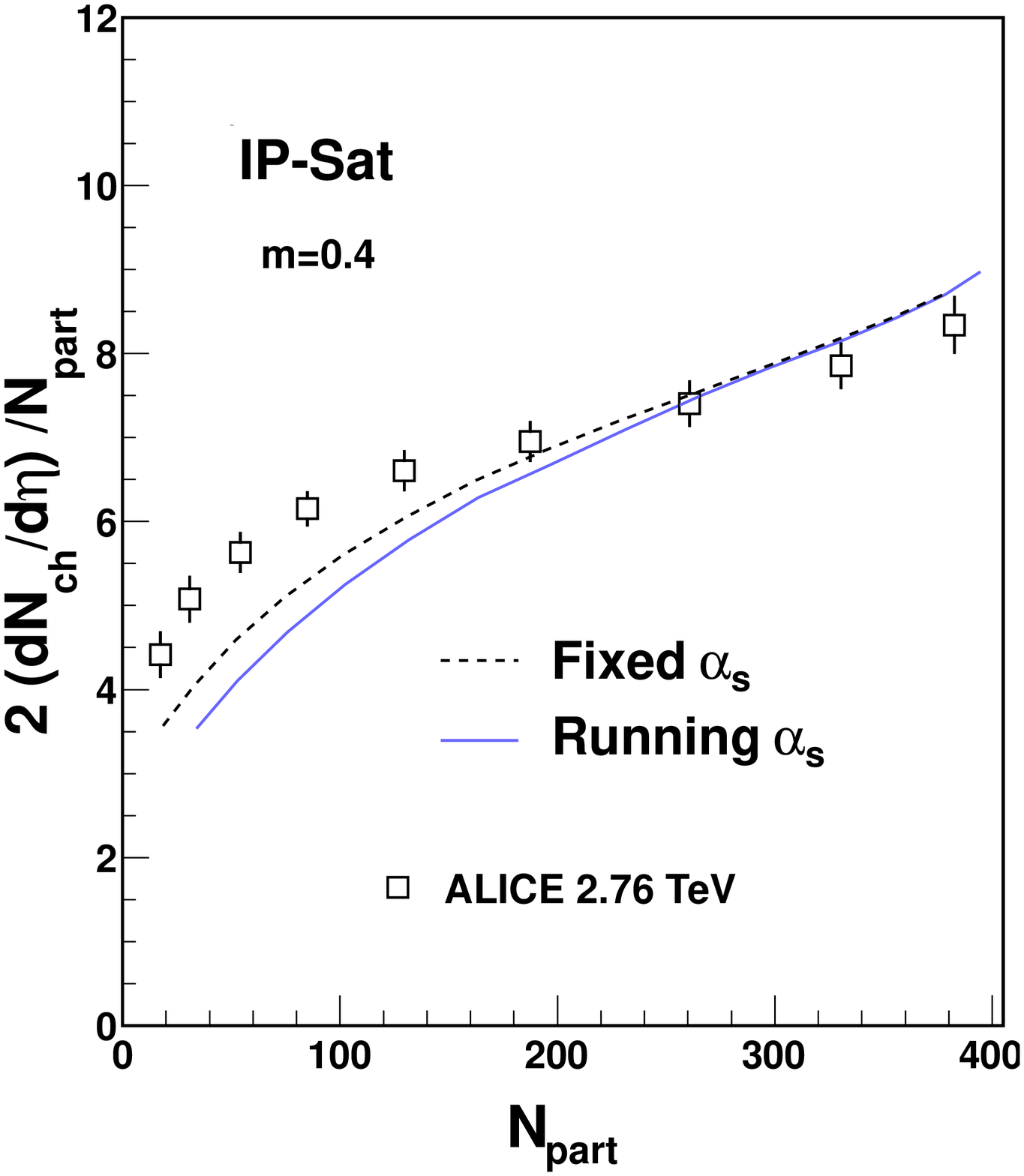}
\caption{Centrality dependance of the inclusive multiplicity in the IP-Sat model compared to RHIC~\cite{nucl-ex/0201005}
 and LHC~\cite{arXiv:1012.1657} data. Left: (fixed coupling) $200$ GeV values for both data and model are multiplied by a factor 2.08. Right: Same plot comparing running (solid curve) and fixed coupling (dashed curve) results in the IP-Sat model.}
\label{fig:cent_ipsat}
\end{figure}

%
We next consider the centrality dependence of the multiplicity at RHIC ($\sqrt{s}=200$ GeV) and LHC ($\sqrt{s}=2.76$ TeV) in the IP-Sat model. While the agreement of the model with data shown in 
fig.~(\ref{fig:cent_ipsat}) is reasonably good for the most central collisions, a systematic deviation is seen for lower centralities, and the model underpredicts the data. While within the range of the theoretical uncertainties outlined thus far, this systematic discrepancy leaves significant room for final state entropy production, which is expected to be more significant for more peripheral collisions. See for instance refs.~\cite{arXiv:1103.1259,{arXiv:1107.5296}} that estimate the amount of entropy production. As the right plot of fig.~(\ref{fig:cent_ipsat}) shows, running coupling effects are less important for the most central collisions but introduce significant uncertainties relative to the fixed coupling results for more peripheral collisions. 

\begin{figure}[h]
\includegraphics[width =7cm, height =7cm]{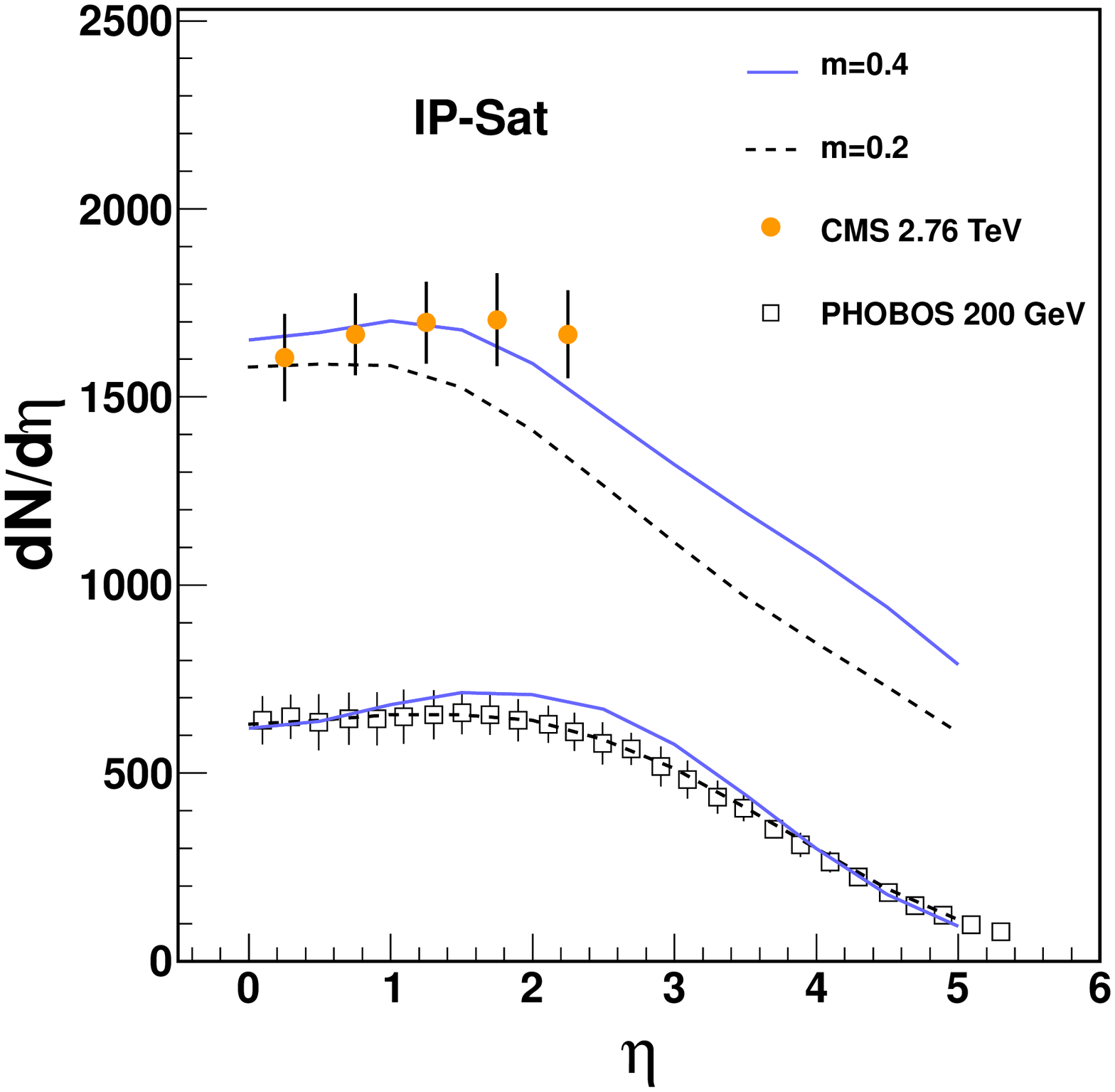}
\includegraphics[width =7cm, height =7cm]{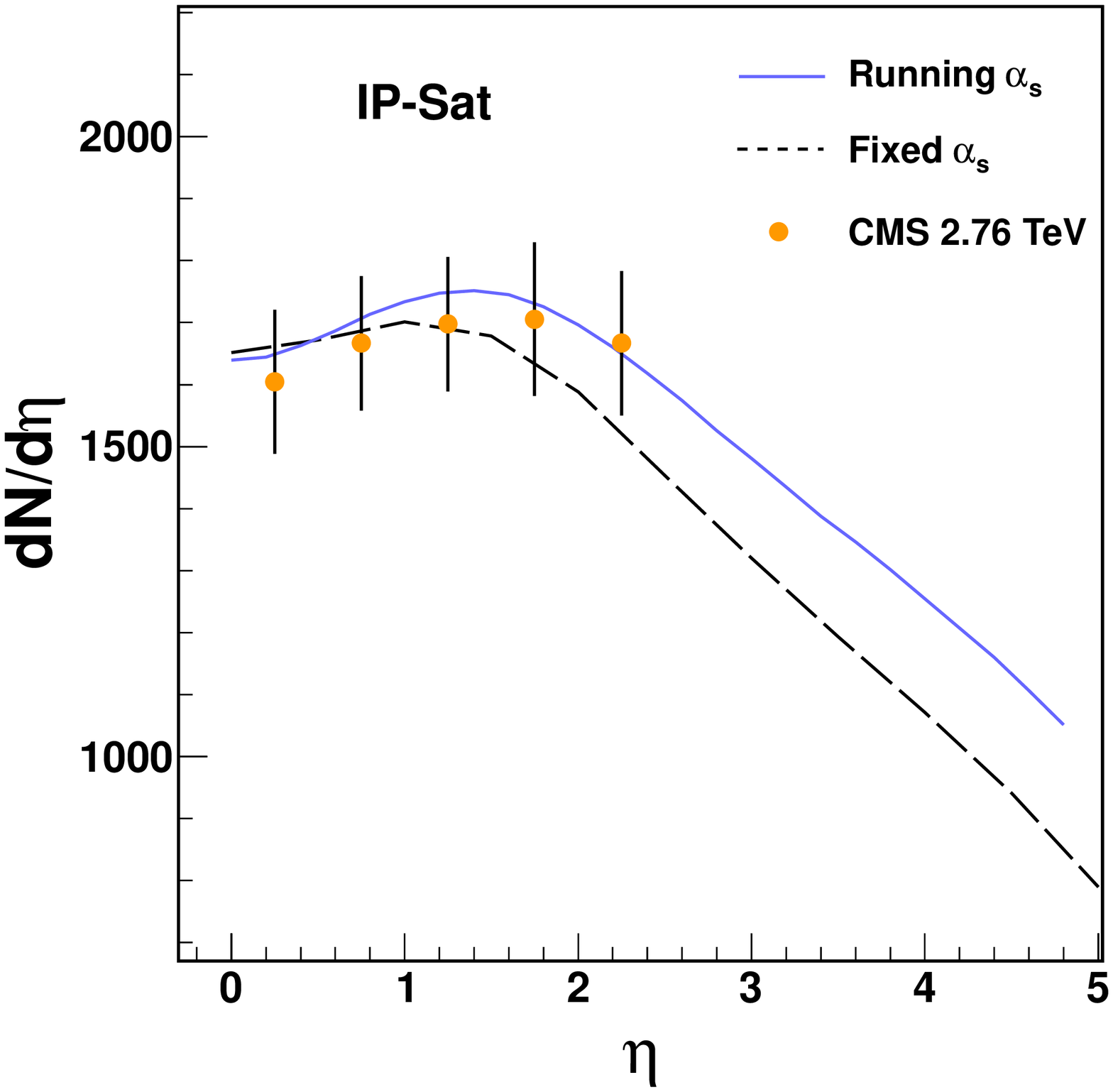}
\caption{Pseudo-rapidity distribution from $\kt$-factorization formula. Left: fixed coupling results for 200 GeV and 2.76 TeV. Right: Same plot(m=0.4) at 2.76 TeV with running (solid curve) and fixed coupling(dashed curve) in $\kt$-factorization formula. Data points are from ref.~\cite{nucl-ex/0210015, arXiv:1107.4800}}
\label{fig:etadist_AA}
\end{figure}

\begin{figure}[h]
\includegraphics[width =8.5cm, height =7cm]{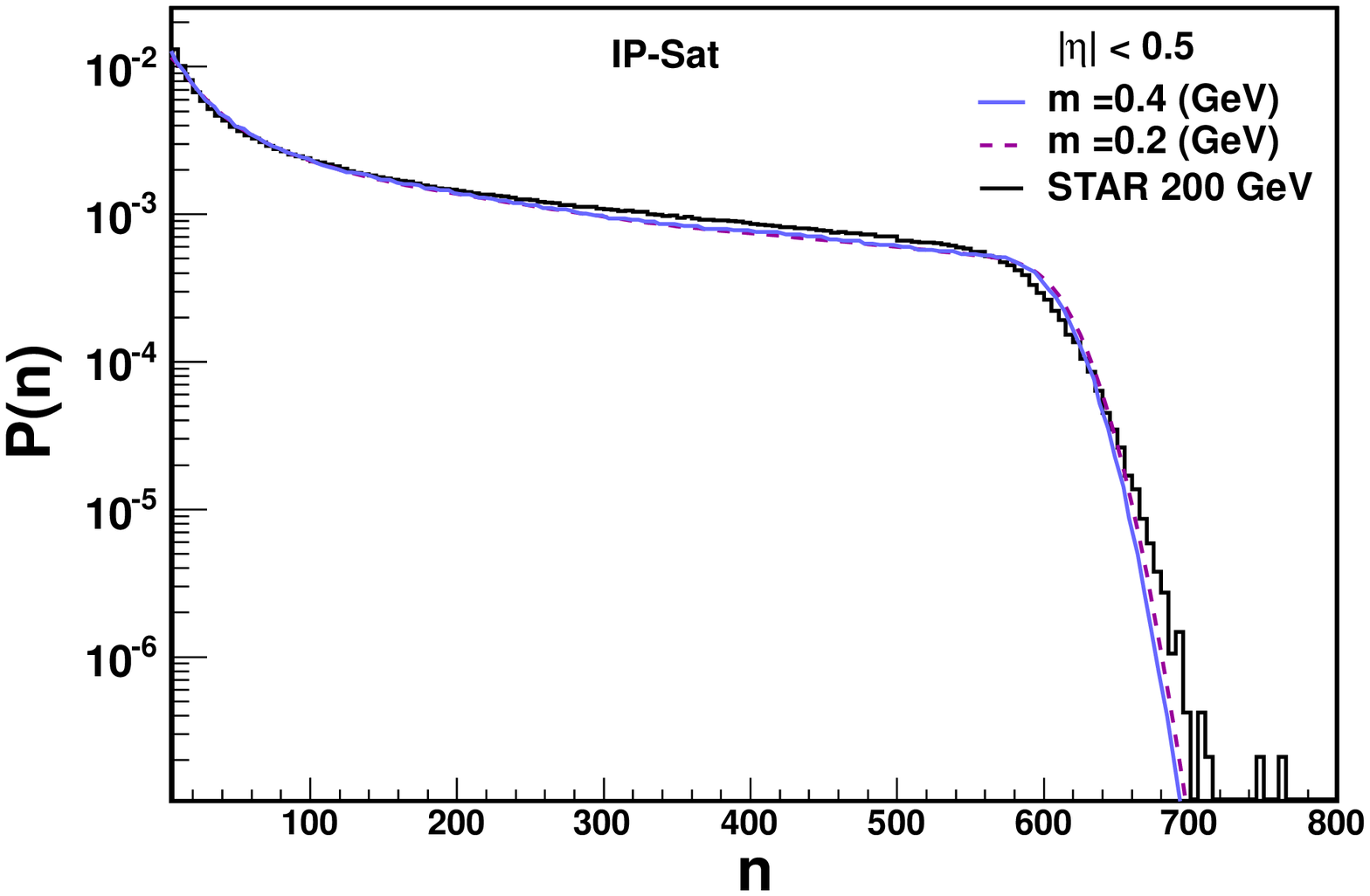}
\includegraphics[width =8.5cm, height =7cm]{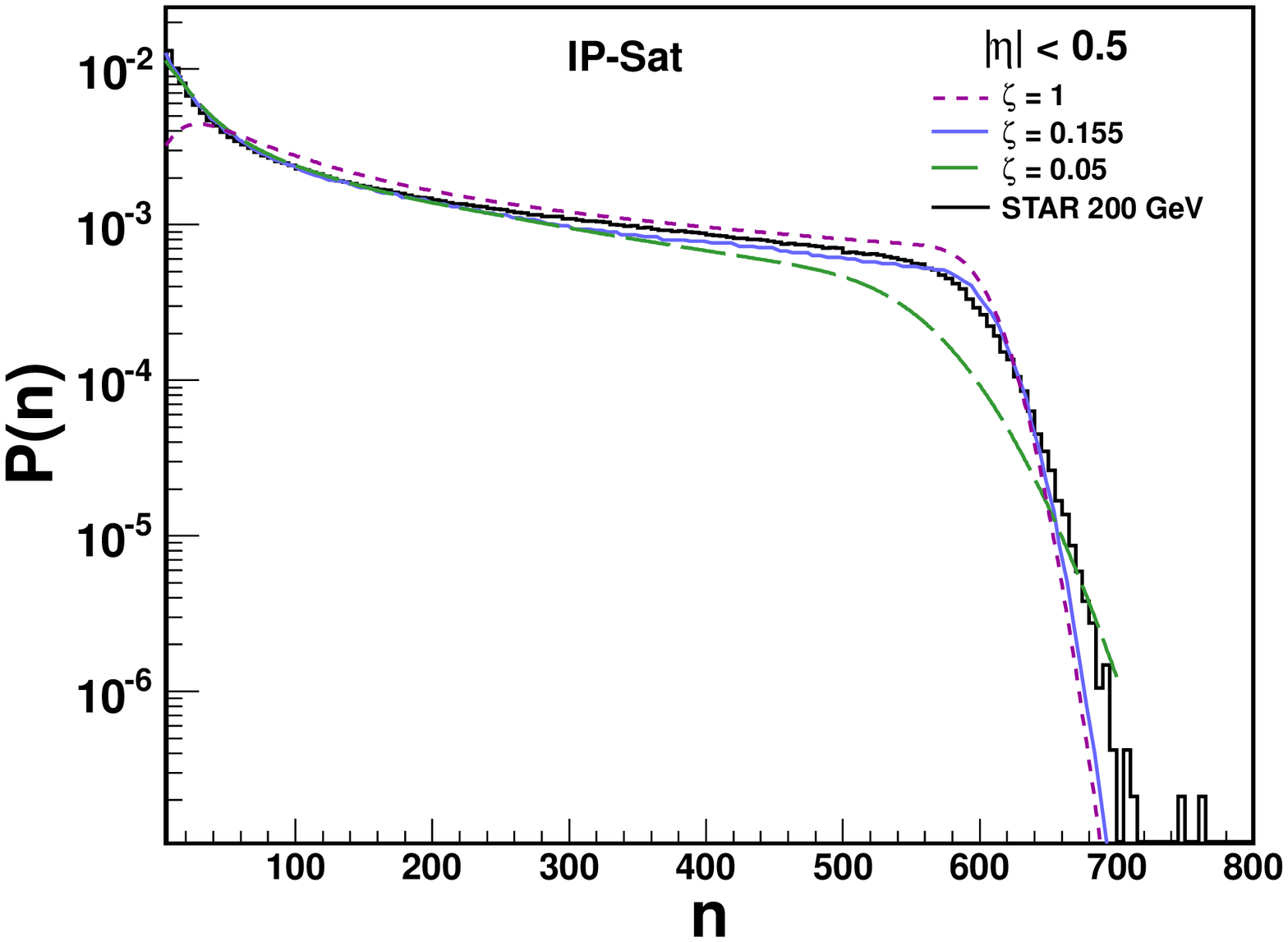}
\caption{Left:Multiplicity distribution for Au+Au collisions in the IP-Sat model compared to uncorrected data(histogram \cite{arXiv:0808.2041}) for different values of $m$. Both data and model plots are normalized for better comparison. Right:Multiplicity distribution for Au+Au collisions at 200 GeV and its sensitivity to the non perturbative constant $\zeta$.}
\label{fig:multdist_m}
\end{figure}

Fig.~(\ref{fig:etadist_AA}) shows the pseudo-rapidity distributions in the IP-Sat model compared to data for Au+Au collisions at 200 GeV (PHOBOS) and  Pb+Pb collisions at 2.76 TeV (ALICE and CMS). Firstly, one sees that the results are sensitive to the infrared cut-off, with improved agreement seen for $m=0.4$ GeV. Further, the rapidity distributions are sensitive to the extrapolation of 
the model to larger $x\geq 0.01$ values. We also note that a significantly better fit to the data at higher energies is obtained by including running coupling effects.  
%
We now employ eq.~(\ref{eq:mult-dist.}) to compute the multiplicity distribution in A+A collisions. While eq.~(\ref{eq:NB}) is computed identically to the p+p case, we need to determine the impact parameter distribution differently from the prescription used for the p+p case in eq.~(\ref{eq:imp-prob-dip}). The expression  
\begin{equation}
{dP_{\rm inel.}\over d^2 \bt} = {1-\left(1-\sigma_{\rm NN} T_{\rm AB}\right)^{AB}  \over \int d^2 \bt \left(1-\left(1-\sigma_{\rm NN} T_{\rm AB}\right)^{AB} \right)} \,,
\label{eq:imp-prob-eik}
\end{equation}
gives a better description of the impact parameter distribution in A+A collisions\footnote{For A+A collisions, the published data~\cite{arXiv:0808.2041} points are uncorrected requiring an additional parameter in contrast to the p+p case. The average multiplicity is 
\begin{equation}
{\bar n}(\bt) = C_{m} \, \left. dN(\bt)\over d\eta \right |_{|\eta|<0.5}\,,
\end{equation}
where the pre-factor $C_{m}$ is the additional parameter specific to the  multiplicity distribution and is tuned to provide a good fit to the uncorrected multiplicity distribution.}. As in the computation of $N_{\rm part}$, $\sigma_{\rm NN} \sim$ 62 (41) mb for 2.76 (0.2) TeV,  is standard and not varied. The saturation scale in this computation is determined at the median value of the impact parameter $\bt^{\rm med.} =15$ GeV$^{-1}$. With these assumptions, the only parameters in computing $P(n)$ are $m$ and $\zeta$, the parameter controlling the width of the multiplicity distributions. In fig.~(\ref{fig:multdist_m}) (left) we see that the multiplicity distributions are insensitive to variations in $m$. Fig.~(\ref{fig:multdist_m}) (right) shows the result of varying $\zeta=0.01$-$1$. Interestingly, we find that the best fit is found for the value of $\zeta=0.155$ that also gives the best fit to the $p+p$ data. 

For the A+A collision data discussed here, we presented results for a) the energy dependence of central (0-6\%) A+A collisions, b) the centrality dependence at two different energies, c) the rapidity dependence (at RHIC for 0-6\% centrality and at LHC for 0-5\% centrality), d) the multiplicity distributions (which are minimum bias).
{Unlike p+p collisions, for A+A collisions, we do not use an energy dependent normalization (because the nuclear size does not grow appreciably with energy unlike the proton)  but a constant factor that is fit to the single inclusive distribution at one energy at $\eta=0$ and used to fit these four data sets a) -d). For the IP-Sat model (the only model considered here), the value of this normalization ($K$-factor) is 0.07 for the mass term m=0.4}

\section{Results for $p$+A collisions}

\begin{figure}[h]
\includegraphics[width =7cm, height =6.5cm]{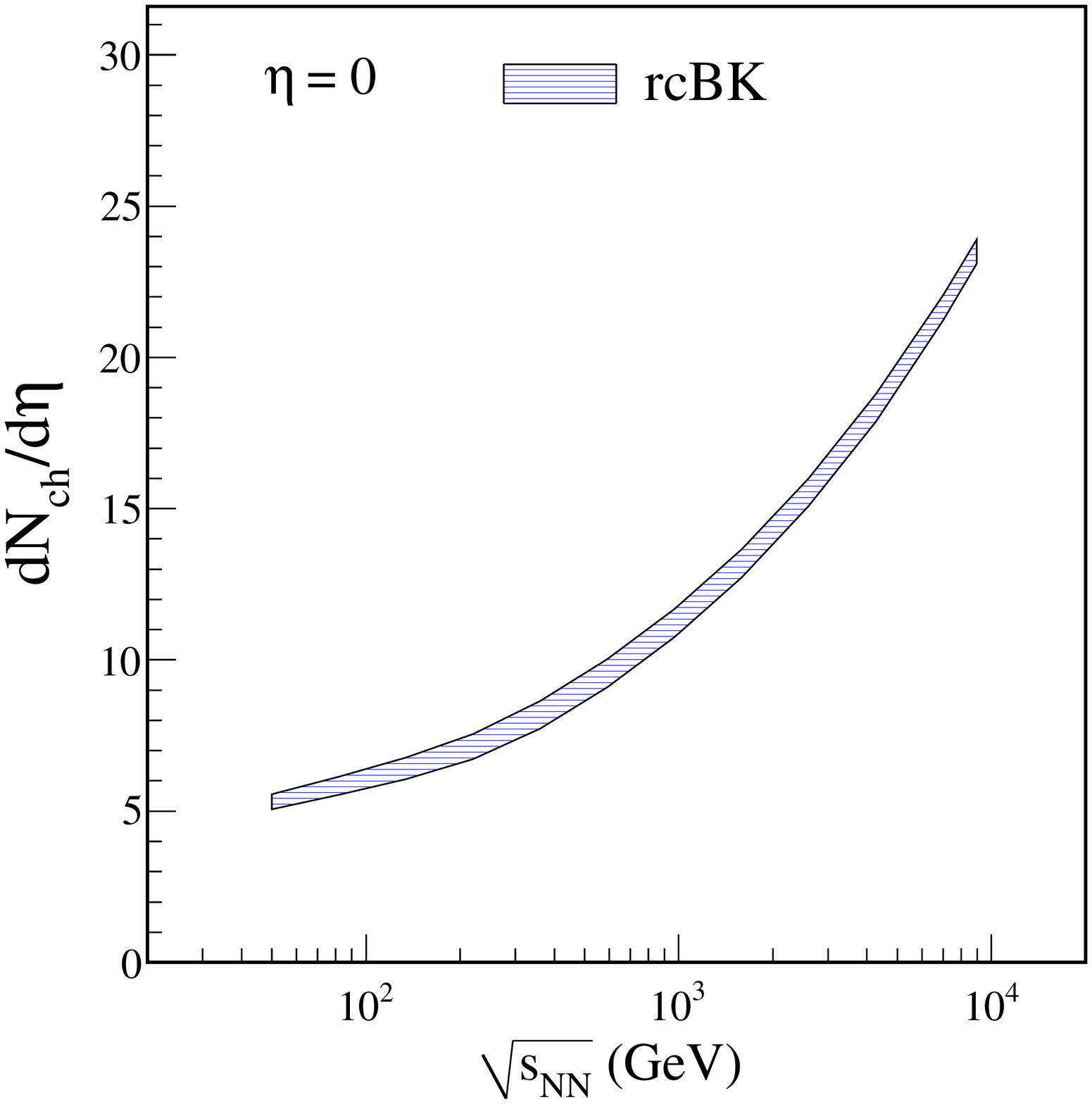}
\includegraphics[width =7cm, height =6.5cm]{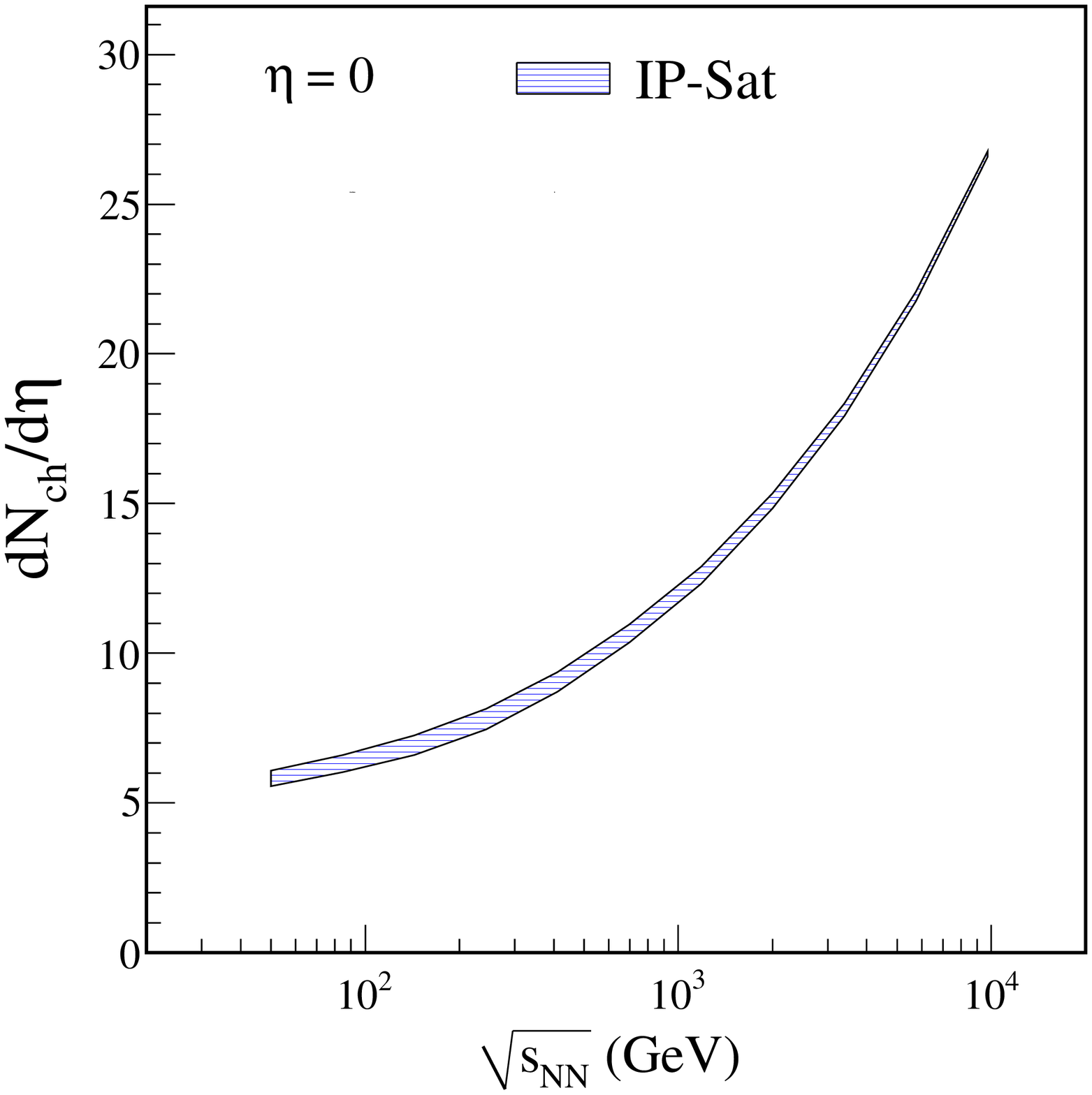}
\caption{Energy dependence of the minimum bias single inclusive multiplicity at $\eta=0$ in p+A collisions from $k_\perp$-factorized unintegrated distributions determined in the rcBK and IP-Sat models. The distribution is normalized with respect to the PHOBOS d+Au data ~\cite{Back:2003hx} for $200$ GeV. The band represents the uncertainty in the calculation due to the variation of the mass term in the range of $0.2$-$0.4$ GeV.}
\label{fig:avgmult_pA}
\end{figure}

\begin{figure}[h]
\includegraphics[width =7cm, height =7cm]{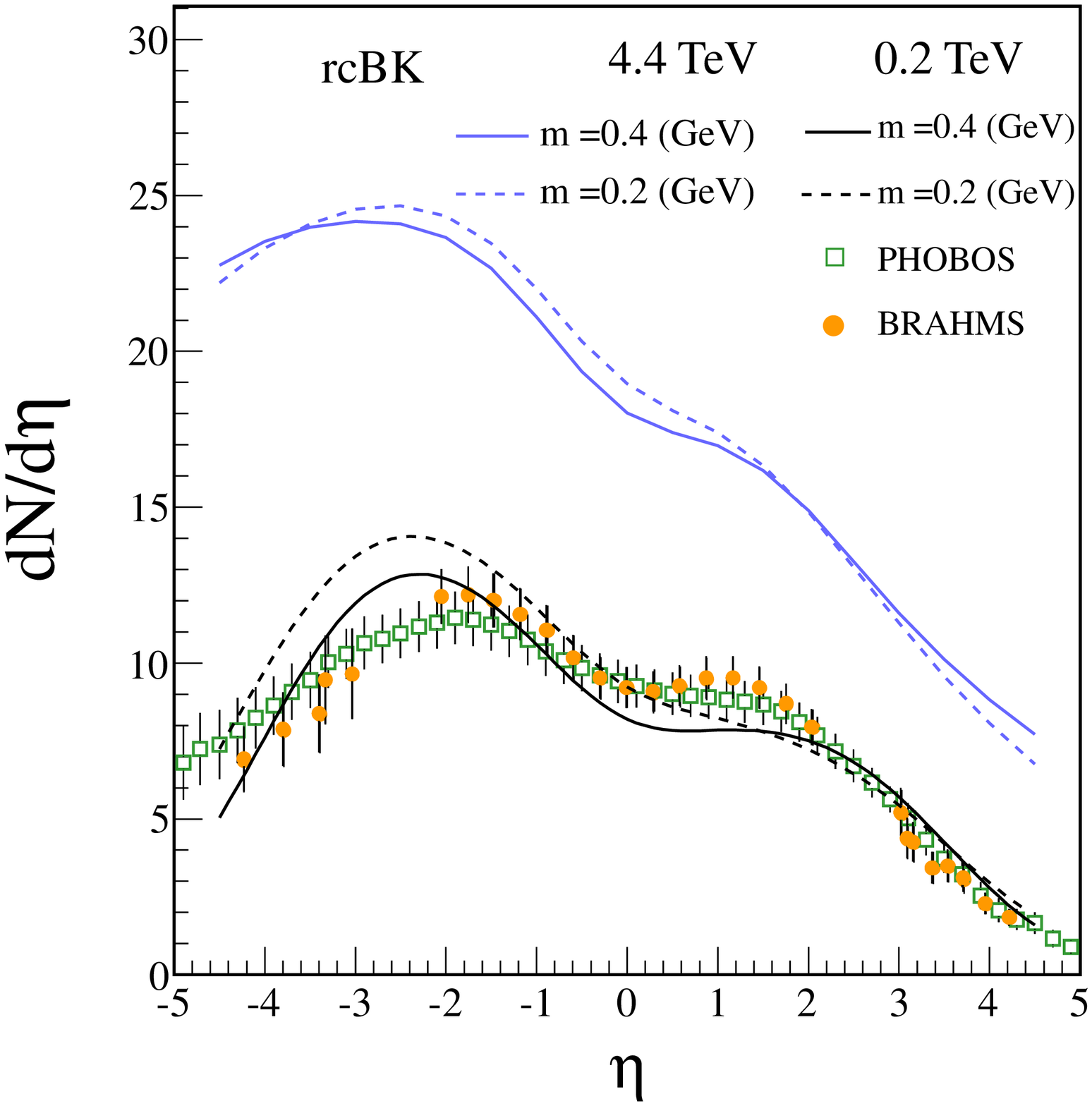}
\includegraphics[width =7cm, height =7cm]{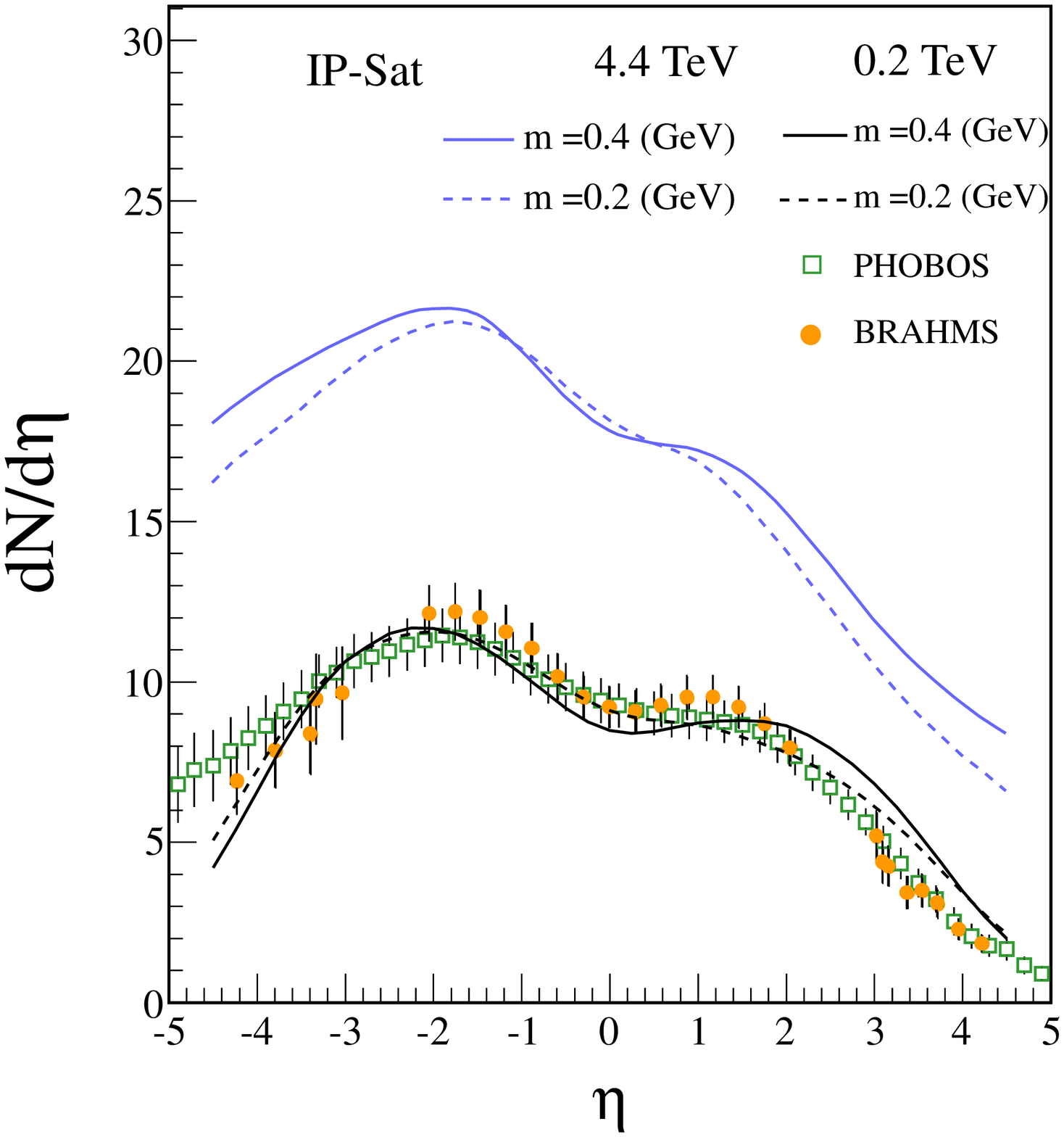}
\caption{Pseudo-rapidity distribution for minimum-bias p+A collision at RHIC and LHC energies. Prediction from rcBK shown for mass term $m$=0.2 and 0.4 GeV. Data points are from ref.~\cite{Back:2003hx, nucl-ex/0401025}}
\label{fig:etadist_pA}
\end{figure}

We computed the min-bias average multiplicity at mid-rapidity  for the IP-Sat and  rc BK models for p+A collisions\footnote{The non-perturbative scales at the initial rapidity for the rc-BK model ~\cite{Tribedy:2010ab} used here are $Q_{s0,p}^2$=0.09 GeV$^2$ and $Q_{s0,A}^2$=0.44 GeV$^2$.}. The data is normalized to the PHOBOS $200$ GeV d+Au data~\cite{Back:2003hx}. The energy dependence of the average multiplicity is shown in fig.~(\ref{fig:avgmult_pA}), the band corresponds to the variation of $m$ in the range of $0.2$-$0.4$ GeV. Both the models give a comparable energy dependence, with the IP-Sat model giving a slightly higher multiplicity at the highest energies. 
The rapidity distributions for RHIC energies in the two models and predictions for LHC energies are shown in fig.~(\ref{fig:etadist_pA}). The models agree with the 
RHIC data with an accuracy of $\approx 10\%$, which is within the theoretical systematic uncertainty, which we shall discuss further shortly. 

\begin{figure}[h]
\includegraphics[width =7cm, height =6.5cm]{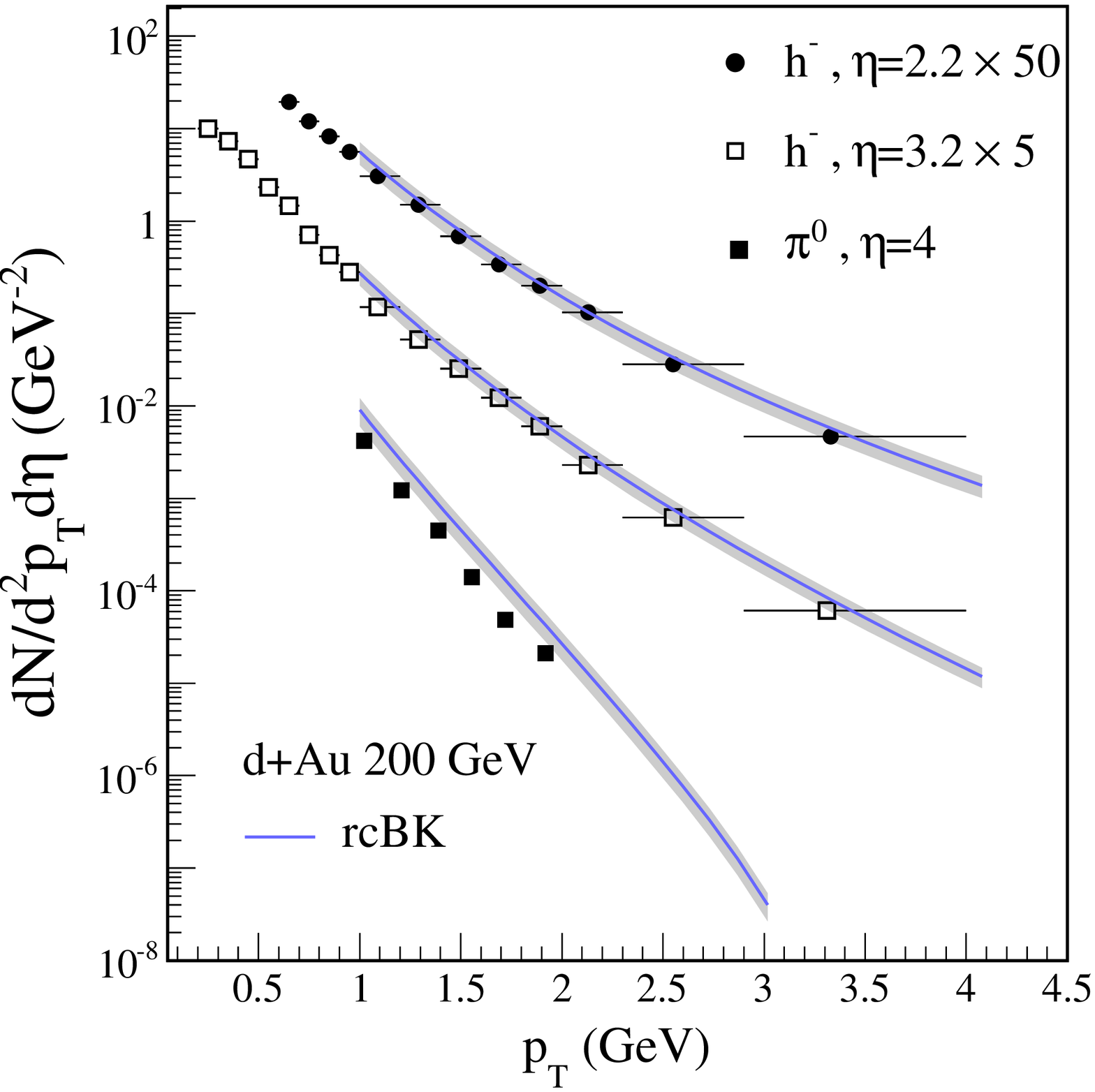}
\includegraphics[width =7cm, height =6.5cm]{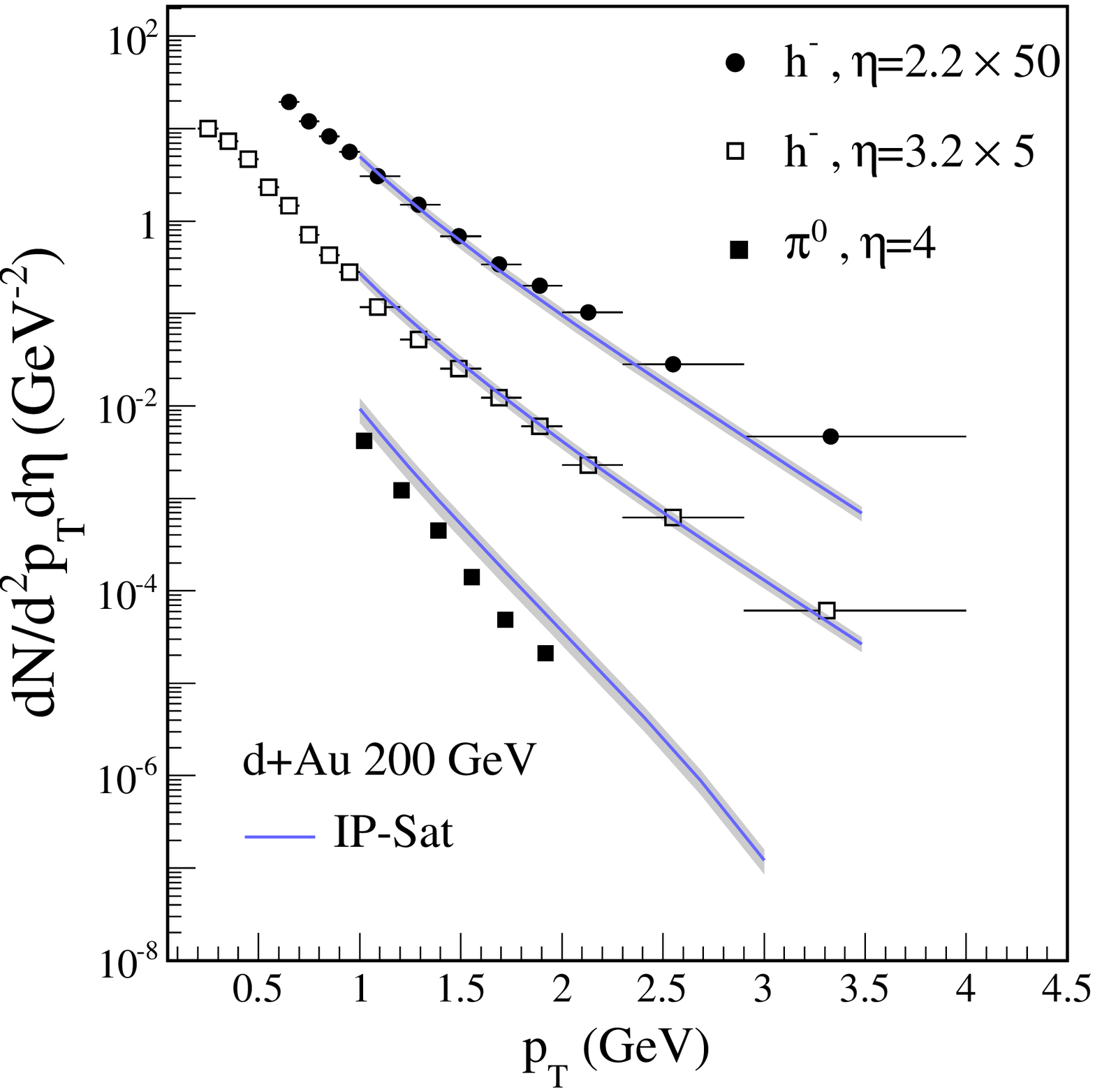}
\caption{Transverse momentum distribution at  forward rapidity  at the highest RHIC energy compared to STAR~\cite{nucl-ex/0602011} and BRAHMS~\cite{nucl-ex/0403005} data. The gray bands show the uncertainty in determination of normalization constant from various sources.}
\label{fig:ptdist_dAu}
\end{figure}

Fig.~(\ref{fig:ptdist_dAu}) shows the  transverse momentum distribution  compared to BRAHMS~\cite{nucl-ex/0403005} and STAR~\cite{nucl-ex/0602011} $200$ GeV data for $h^{-}$ and $\pi^0$ at forward rapidities. 
For the $h^-$ case, we have included 15$\%$ isospin correction  in the normalization constant, to be discussed further below.  Predictions for the $p+A$ transverse momentum distributions for charged hadrons at $\eta=0$ for LHC energies with this fixed normalization are shown in fig.~(\ref{fig:ptdist_pA}). 

\begin{figure}[h]
\includegraphics[width =7cm, height =6.5cm]{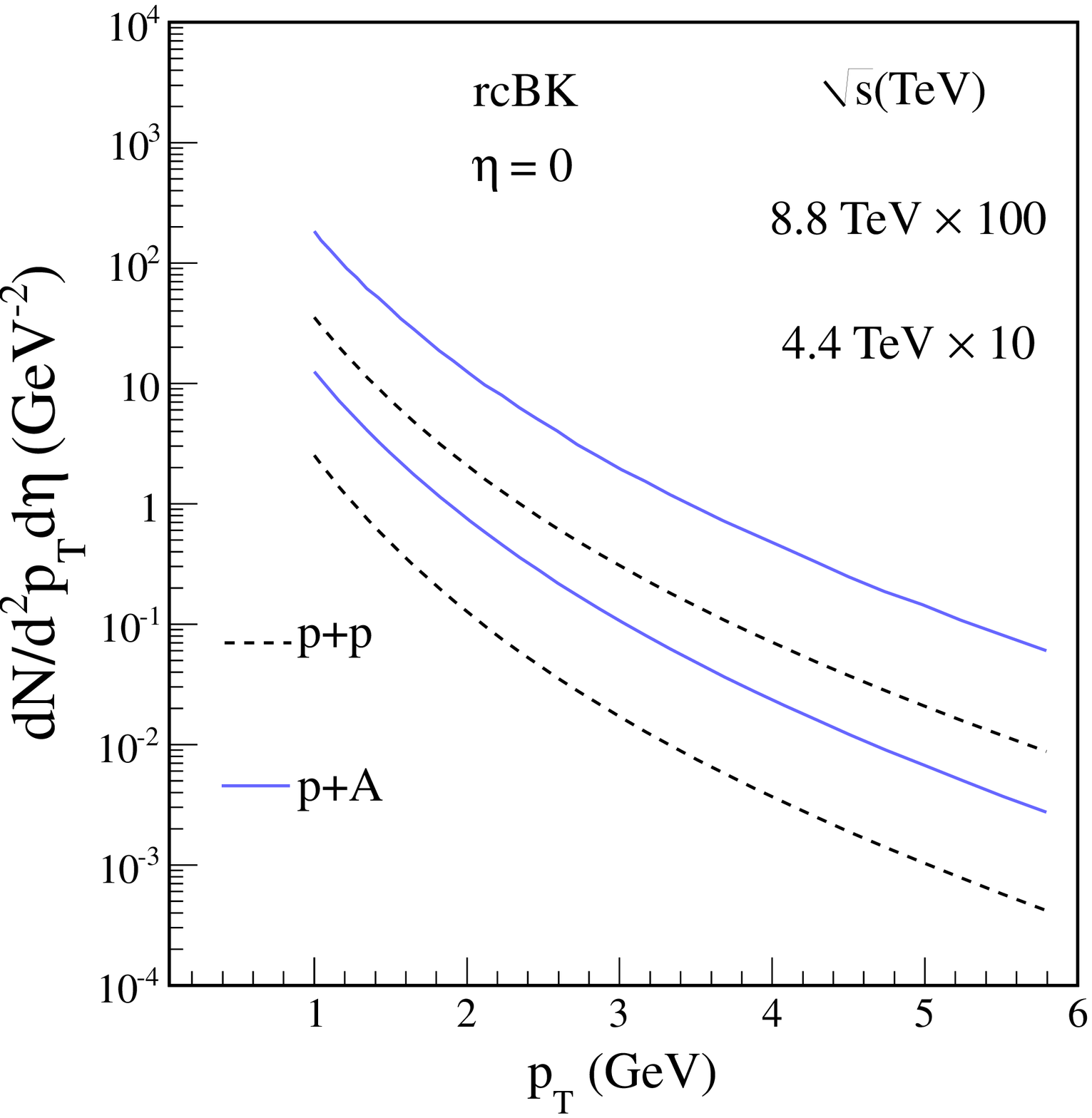}
\includegraphics[width =7cm, height =6.5cm]{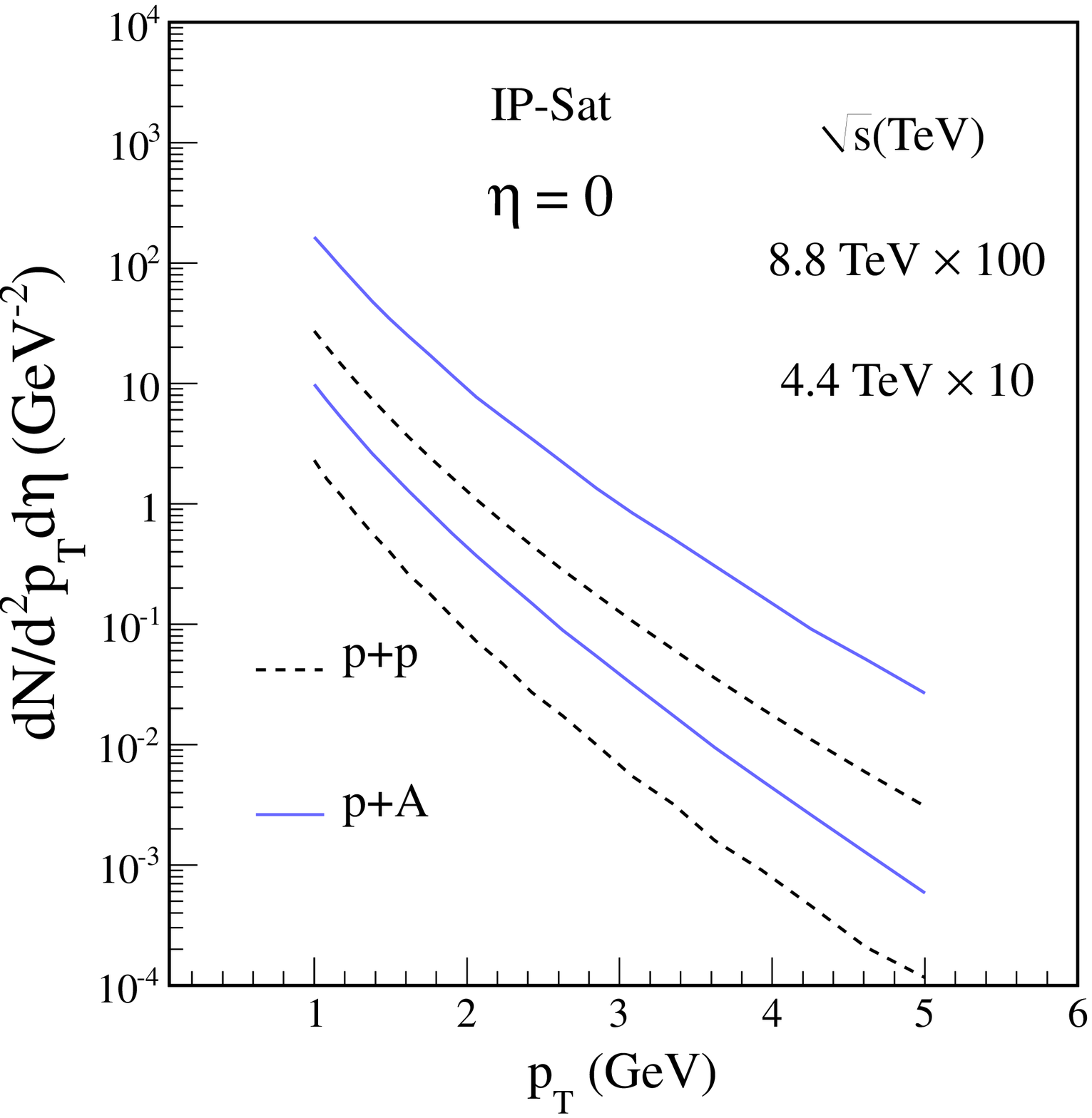}
\caption{Transverse momentum distribution at mid-rapidity for minimum-bias p+p and p+A collisions..}
\label{fig:ptdist_pA}
\end{figure}

\begin{figure}[h]
\includegraphics[width =7cm, height =6.5cm]{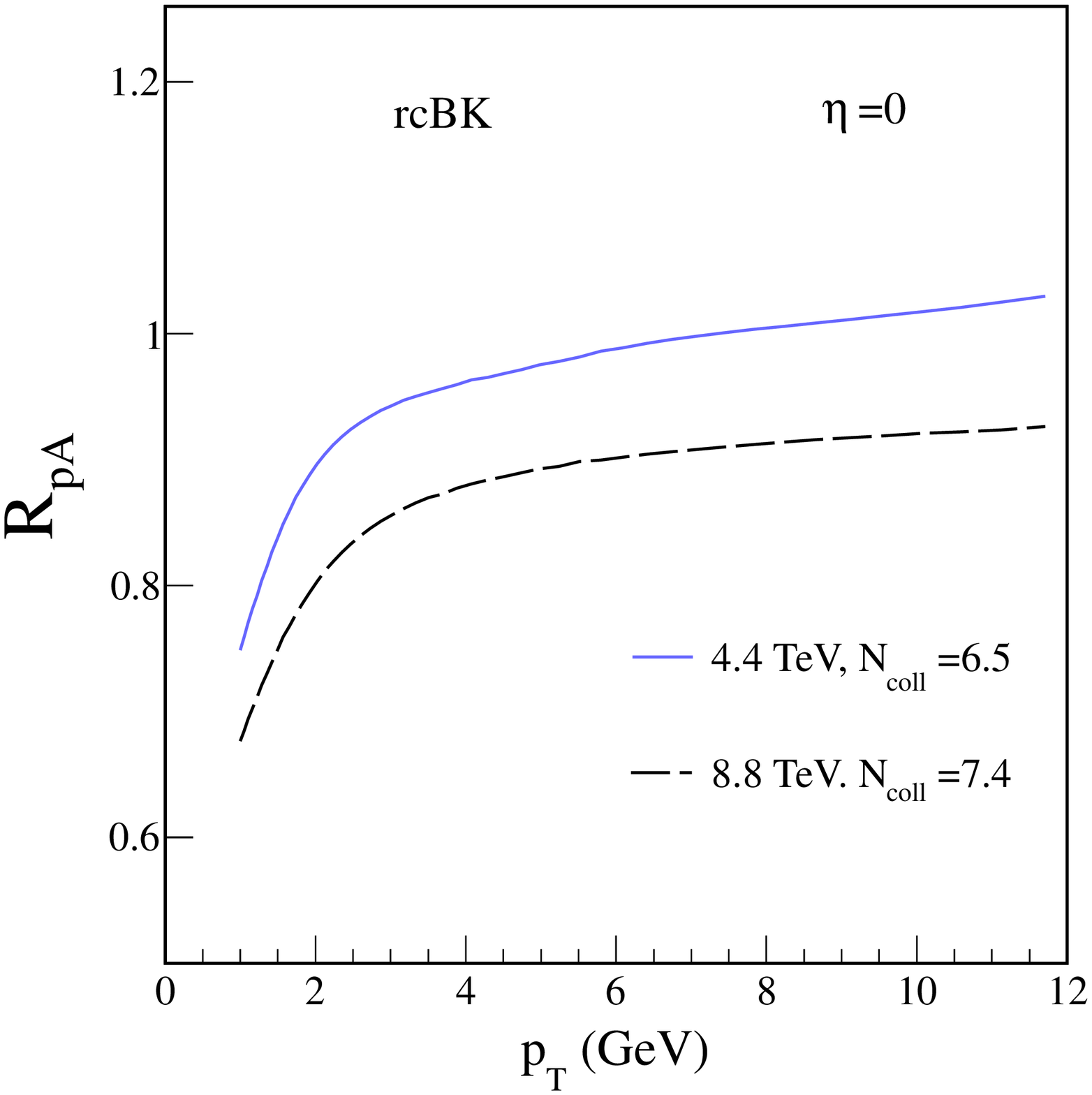}
\includegraphics[width =7cm, height =6.5cm]{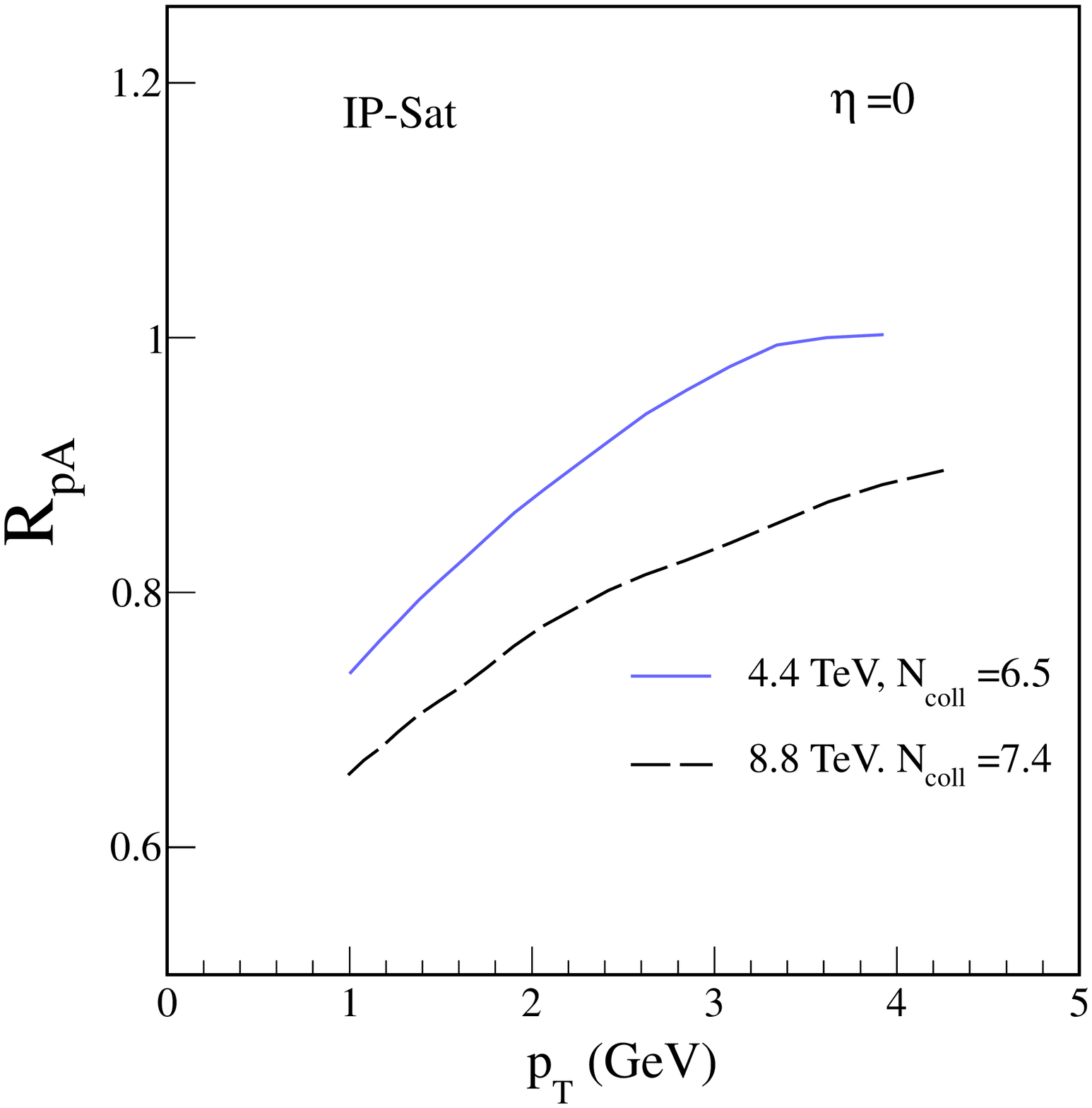}
\caption{
Nuclear modification factor for charged hadrons at mid-rapidity  for p+A collisions at LHC energies.}
\label{fig:ptdist_pA}
\end{figure}
We compute the nuclear modification factor anticipated in $p+A$ collisions at the LHC in the IP-Sat and rcBK models. $R_{pA}$ in both models, for $\sqrt{s}=4.4$ TeV/nucleon, approaches unity at $p_\perp\sim 5$ GeV. For $\sqrt{s}=8.8$ TeV/nucleon, the suppression persists to higher $p_\perp$. The slope of $R_{pA}$ however appears quite different in the two models. 

Regarding the overall normalization, for d+Au we have data available at only one energy. As noted, data and predictions include a) the energy dependence, b) the rapidity dependence at RHIC and LHC, and the the single inclusive pt distributions. 

{For the IP-Sat model the form of the normalization is  $A/(\pi b_{\rm max}^2)$  where, $A$ = 0.25 with mass term $m=0.4$ GeV. Unlike $p+p$ here $b_{\rm max} = 9.5$ fm is a fixed number which is the maximum range of impact parameter to obtain minimum bias distribution and does not change with energy. For the rc-BK model a single normalization constant $A$ = 0.032 (with $m=0.4$ GeV) absorbs 
\footnote{eq.\ref{eq:ktfact1} for min-bias p+A collision the rc-BK model normalization constant includes the pre factor ($\frac{4 \alpha_S}{\pi C_F (2\pi)^{5}} \frac{S_{p,A}}{(\pi R_p^2) (\pi R_A^2)}$);  here $R_p$, $R_A$ corresponds  to the radii of the proton and nucleus and $S_{p,A}$ is the overlap area.}
all the constant pre factors of eq.\ref{eq:ktfact1}. 
These normalization constants are obtained from a fit to the PHOBOS pseudo rapidity distribution;  one obtains an $\sim 8\%$ higher value of $A$ when the fit is performed only to the data points at $\eta=0$. Using the BRAHMS data for normalization also gives higher values for $A$ which, along with other numerical uncertainties, contributes to the bands shown in fig.~\ref{fig:ptdist_dAu}.
The significant difference in $A$ for IP-Sat and rcBK for d+Au collisions is because the the area of overlap and other terms in the pre-factor of kt-factorization are absorbed in $A$ for rc-BK and cannot be separated from the ``$K$ factor".}
 For the IP-Sat model, eq.~(\ref{eq:unint-gluon})  includes those factors and $A$ is of the order of 1. This apparent difference in the two models doesn't affect any of our final results since same normalization is consistently used everywhere. Note also that in conversion from d+Au to p+A numbers we have used an additional factor of 1.6/2 in the normalization, which is standard in such conversions in the literature. In computing $R_{pA}$, the result depends on $N_{\rm coll}$, which is sensitive to the proton inelastic cross-section. Since $b_{\rm max,proton}$ grows with energy, one finds for the energies $\sqrt{s}=4.4,8.8$ TeV that ratio $b_{\rm max,proton}^2(8.8 {\rm TeV}) / b_{\rm max,proton}^2(4.4 {\rm TeV}) \sim N_{\rm coll}(8.8 {\rm TeV})/ N_{\rm coll}(4.4 {\rm TeV})$, a result  consistent with expectations of $N_{\rm coll}$ from Glauber 
approaches~\cite{d'Enterria:2010hd}; numbers quoted are in agreement with our ratio to 5\%.

\section{Caveats and comparisons}

We would be remiss not to discuss the known sources of theoretical uncertainties in our framework; a brief discussion was also presented in ref.~\cite{Tribedy:2010ab}. One that we alluded to previously is the $k_\perp$ factorization framework. Corrections to these come both from additional multiple scattering contributions in the dense projectile-dense target limit and from higher order contributions~\cite{arXiv:1009.0545,arXiv:1112.1061}; We estimate differences with the classical Yang-Mills multiple scattering effects to be no larger than other theoretical uncertainties and is accounted for an approximately $15\%$ difference in normalization between the integrated multiplicity and the $p_\perp$ distributions. The higher order contributions are constrained by a normalization constant that's fit at a given energy. We have also shown in fig.~(\ref{fig:etadist_AA}) the effect of running coupling on multiplicity distributions. Another source of theoretical uncertainty arises from logarithmic sensitivity to the infrared cut-off seen in the single inclusive multiplicity (but not in the multiplicity distribution); the additional (classical Yang-Mills) scattering effect we alluded to also makes the distributions infrared finite. Even though the Yang-MIlls contribution is understood, it is at present cumbersome to include it in a ``global" analysis. 

In the IP-Sat model, as discussed previously in ref.~\cite{Tribedy:2010ab}, there is a sensitivity to the different data sets for the dipole cross-section that gave best fits to the then extant HERA data. We noted in ref.~\cite{Tribedy:2010ab} that this sensitivity decreased with increasing energy. The sensitivity to the different fits to the gluon distribution is particularly severe at large $p_\perp$ where small differences are accentuated because of the steeply dropping cross-section. At high $p_\perp$ and large rapidities, the unintegrated distributions are sensitive to $x\geq x_0 (=0.01)$. Guided by quark counting rule prescriptions, we choose $\phi (x >  x_0) = \phi (x_0) (1-x)^4/(1-x_0)^4$, as in eq.~(20) of ref.~\cite{Tribedy:2010ab}, except with the parameter $\lambda$ there set to zero. Further, as noted previously, the IP-Sat fits in ref.~\cite{KMW} were performed prior to the combined ZEUS-H1 data . One expects a revised fit to have an impact on the numbers extracted in ref.~\cite{KMW}. Finally, there is theoretical uncertainty arising from our imperfect knowledge of fragmentation functions. 

Given the many uncertainties, one has to think of these predictions at this stage as  having considerable elasticity of $20\%$-$50\%$ variability depending on the quantity studied. Only qualitative differences in measured distributions from theoretical projections (such as  $R_{pA} > 1$ at low $p_\perp$ at the LHC) can rule out models at present. Else, one has to check whether variations in parameters without adding new ones can accommodate the data. This is no different in spirit from global fits in perturbative QCD. Given that we have not tried to fine tune parameters, the agreement with the data presented here (in combination to the 
results in ref.~\cite{Tribedy:2010ab}) is quite good considering that the dipole approach also gives a good description of small $x$ HERA data. 

We will now briefly compare our results with some recent related approaches. In ref.~\cite{arXiv:1111.2312}, results for the pseudo-rapidity distributions in p+p and A+A collisions at RHIC and the LHC are presented (based on previous work in ref.~\cite{arXiv:1007.2430}) and predictions made for p+A collisions. The results shown are primarily for the b-CGC model which we also considered previously. While the b-CGC model does well for the pseudo-rapidity distributions, it does less well relative to the IP-Sat model for the multiplicity distribution $P(n)$; as fig.~(\ref{fig:multdist_pp}) indicates, the latter works quite well. When it comes to nuclei, ref.~\cite{arXiv:1111.2312} assumes that an additional gluon cascade mechanism to explain the faster energy dependence of the charged particle multiplicity. However, as pointed out in ref.~\cite{arXiv:1104.3725}, the difference in the energy dependence of the charged particle multiplicity in A+A collisions relative to p+p collisions is economically explained by the fact that the larger saturation scales in nuclei lead to more phase space for bremsstrahlung; while this may be conceptually similar to the cascade posited in ref.~\cite{arXiv:1111.2312}, no further modification of the formalism is necessary in the former approach. Indeed, we see in fig.~(\ref{fig:avgmult_AA}), that a good agreement is obtained for both p+p and A+A data in the IP-Sat model. A similar quality of agreement\footnote{Small differences in the 
predictions there for pA pseudo-rapidity distributions with those presented here can be attributed primarily to slight different initial values for the saturation scale.} is seen in the rcBK model (combining results in refs.~\cite{Tribedy:2010ab} and ~\cite{arXiv:1011.5161}). In ref.~\cite{arXiv:1111.3031}, results in the KLN model for the pseudo-rapidity distributions in p+p, d+A and A+A collisions at RHIC are presented, as are results for the LHC in p+p and A+A collisions, and predictions are made for the same in p+A collisions at the LHC. Similarly, results are presented for the $P(n)$ in p+p collisions and predictions of the same made for p+A collisions. The KLN model gives good results for the pseudo-rapidity distributions at RHIC energies and in p+p and A+A distributions at LHC energies, which are similar to the models considered here. While the multiplicity distribution $P(n)$ is motivated similarly to our discussion, the expression in ref.~\cite{arXiv:1111.3031} does not depend on impact parameter; it will be interesting to see if a similar quality of agreement is obtained for higher $n$ as in our fig.~(\ref{fig:multdist_pp}). The KLN model gives a better agreement to the centrality dependence of the charged particle multiplicity when compared to fig.~(\ref{fig:cent_ipsat}); our result however leaves open the possibility that there may be significant entropy generation in the final state in more peripheral collisions. We also note that the KLN model does not at present allow for a global fit because it is not constrained by HERA data on e+p collisions~\cite{arXiv:1005.0631}.

Multiplicity distributions are sensitive only to $p_\perp < 1$ GeV. For $p_\perp$ distributions, we first compare our results for $R_{pA}$ to those in ref.~\cite{arXiv:1001.1378}. Only results for $8.8$ TeV/nucleon are presented there, and for forward rapidities. Even so, we see that the suppression is considerably less in our framework. A significant fact is that in $R_{pA}$, one has to account for a growth in the proton minimum bias cross-section with energy, which tends to suppress the denominator and lessen the suppression from what it would be otherwise. We found this effect to be quite significant and modify plots from those similar to ref.~\cite{arXiv:1001.1378} to what we present here. We expect the same effect to impact the result (in fig.~(5) of ref.~\cite{arXiv:1110.2810} using a ``hybrid" formalism~\cite{hep-ph/0506308} corrected by the inelastic contributions introduced in ref.~\cite{arXiv:1102.5327}. The ``hybrid" approach may be more suitable than our approach when distributions are sensitive to large
$x > 0.01$ and high $p_\perp$; significant recent progress~\cite{arXiv:1112.1061} in that approach will allow us to compute systematically corrections beyond the $k_\perp$ factorization framework. Finally, we observe that $R_{pA}$ in the IP-Sat model approaches 
unity more rapidly for $\sqrt{s}=4.4$ TeV/nucleon than for the leading twist shadowing evolved EPS08 curve~\cite{arXiv:1002.2537, arXiv:1105.3919}  plotted in ref.~\cite{arXiv:1111.3646}, while the rcBK curve has a similar behavior to the EPS08 curve. This is not too surprising because shadowing in the IP-Sat model is from higher twist contributions that go away quickly at large $Q^2$; in contrast, there is a significant leading twist window in the rcBK model which may explain the qualitative similarity of the two results in this kinematic window.

\section*{Acknowledgements}
R.V is supported by the US Department of Energy under DOE Contract No.DE-AC02-98CH10886. We thank Subhasis Chattopadhyay and Adrian Dumitru for very valuable conversations. {We thank T. Lappi for bringing a typo in the earlier version of this paper to our attention.}


\begin{thebibliography}{50}

\bibitem{Tribedy:2010ab}
  P.~Tribedy, R.~Venugopalan,
  Nucl.\ Phys.\  {\bf A850}, 136-156 (2011).

\bibitem{MV}L.D. McLerran, R. Venugopalan,  Phys. Rev. {\bf D} {\bf 49}, 2233 (1994); {\it ibid.} {\bf 49}, 3352 (1994); {\it ibid.} {\bf 50}, 2225 (1994).

\bibitem{CGC-reviews}E. Iancu, R. Venugopalan, Quark Gluon Plasma 3, Eds. R.C. Hwa, X.N. Wang,
  World Scientific, hep-ph/0303204; H.~Weigert,
  Prog.\ Part.\ Nucl.\ Phys.\  {\bf 55}, 461 (2005); F. Gelis, E. Iancu, J. Jalilian-Marian, R. Venugopalan, arXiv:1002.0333.

\bibitem{GLR}L.V. Gribov, E.M. Levin, M.G. Ryskin, Phys. Rept. {\bf 100}, 1 (1983); A.H. Mueller, J-W. Qiu, Nucl. Phys. {\bf B} {\bf 268}, 427 (1986).

\bibitem{JIMWLK} Jalilian-Marian, A. Kovner, A. Leonidov, H. Weigert, Nucl. Phys. {\bf B}
  {\bf 504}, 415 (1997); {\it ibid.}, Phys. Rev. {\bf D}
  {\bf 59}, 014014 (1999); E. Iancu, A. Leonidov, L.D. McLerran, Nucl. Phys. {\bf A} {\bf 692}, 583
  (2001); E. Ferreiro, E. Iancu, A. Leonidov, L.D. McLerran, Nucl. Phys. {\bf A} {\bf
  703}, 489 (2002).

\bibitem{BK}I. Balitsky, Nucl. Phys. {\bf B} {\bf 463}, 99 (1996); Yu.V. Kovchegov, Phys. Rev. {\bf D} {\bf 61}, 074018 (2000).

\bibitem{KT} H. Kowalski, D. Teaney, Phys. Rev. D 68, 114005
(2003).

\bibitem{IIM}E.~Iancu, K.~Itakura, S.~Munier,
  Phys.\ Lett.\  B {\bf 590}, 199 (2004).

\bibitem{KMW}H.~Kowalski, L.~Motyka, G.~Watt,
  Phys.\ Rev.\  D {\bf 74}, 074016 (2006).

\bibitem{KW}G.~Watt, H.~Kowalski,
  Phys.\ Rev.\  D {\bf 78}, 014016 (2008).

\bibitem{Albacete}J.~L.~Albacete, Y.~V.~Kovchegov,
  Phys.\ Rev.\  D {\bf 75}, 125021 (2007).

\bibitem{arXiv:1012.4408} 
  J.~L.~Albacete, N.~Armesto, J.~G.~Milhano, P.~Quiroga-Arias and C.~A.~Salgado,
  Eur.\ Phys.\ J.\ C\ {\bf 71}, 1705  (2011).
  
  \bibitem{arXiv:0911.0884} 
  F.~D.~Aaron {\it et al.} [H1 and ZEUS Collaboration],
  JHEP\ {\bf 1001}, 109  (2010).
  
  \bibitem{ZEUS}S. Chekanov et al., [ZEUS Collaboration] Eur. Phys. J. C 21 (2001) 443.

\bibitem{BGV1} J.P. Blaizot, F. Gelis, R. Venugopalan, Nucl. Phys. A 743,57 (2004).

\bibitem{Braun}M.A. Braun, Phys. Lett. {\bf B} {\bf 483}, 105 (2000). 

\bibitem{GelisSV}F.~Gelis, A.~M.~Stasto, R.~Venugopalan,
  Eur.\ Phys.\ J.\  C {\bf 48}, 489 (2006).
  
  \bibitem{hep-ph/9909203} 
  A.~Krasnitz and R.~Venugopalan,
  Phys.\ Rev.\ Lett.\ \ {\bf 84}, 4309  (2000).
  
  \bibitem{hep-ph/0305112} 
  A.~Krasnitz, Y.~Nara and R.~Venugopalan,
  Nucl.\ Phys.\ A\ {\bf 727}, 427  (2003).
  
  \bibitem{hep-ph/0303076} 
  T.~Lappi,
  Phys.\ Rev.\ C\ {\bf 67}, 054903  (2003).
  
  \bibitem{arXiv:1005.0955} 
  J.~-P.~Blaizot, T.~Lappi and Y.~Mehtar-Tani,
  Nucl.\ Phys.\ A\ {\bf 846}, 63  (2010).

\bibitem{DumitruGMV}{A. Dumitru, F. Gelis, L. McLerran, R. Venugopalan}, Nucl. Phys. {\bf A} {\bf
  810}, 91 (2008).

\bibitem{GelisLM}F. Gelis, T. Lappi,  L. McLerran, Nucl. Phys. A828 (2009) 149. 

\bibitem{hep-ph/0105268} 
  A.~Dumitru and L.~D.~McLerran,
  Nucl.\ Phys.\ A\ {\bf 700}, 492  (2002).

\bibitem{arXiv:1101.5922} 
  P.~Tribedy and R.~Venugopalan,
  arXiv:1101.5922 [hep-ph].
  
\bibitem{LappiSV}{T. Lappi, S. Srednyak, R. Venugopalan}, JHEP {\bf 1001} 066 (2010).

\bibitem{Kovner:2011pe} 
  A.~Kovner and M.~Lublinsky,
  arXiv:1109.0347 [hep-ph].

\bibitem{Dumitru-etal}A.~Dumitru, K.~Dusling, F.~Gelis, J.~Jalilian-Marian, T.~Lappi and R.~Venugopalan,
  Phys.\ Lett.\ B\ {\bf 697}, 21  (2011).
  
  \bibitem{Dusling:2012ig} 
  K.~Dusling and R.~Venugopalan,
  arXiv:1201.2658 [hep-ph].

\bibitem{CMS-ridge}V.~Khachatryan {\it et al.} [CMS Collaboration],
  JHEP\ {\bf 1009}, 091  (2010).
  
\bibitem{ua53} UA5 Collaboration, Z. Phys. C43 (1989) 357.

\bibitem{alice2} ALICE Collaboration, Eur.Phys.J.C68:345-354,2010.

\bibitem{Khachatryan:2010nk}
  V.~Khachatryan {\it et al.}  [CMS Collaboration],
  JHEP {\bf 1101}, 079 (2011).

  
  \bibitem{nucl-ex/0602011} 
  J.~Adams {\it et al.} [STAR Collaboration],
  Phys.\ Rev.\ Lett.\ \ {\bf 97}, 152302  (2006).

\bibitem{nucl-ex/0403005} 
  I.~Arsene {\it et al.} [BRAHMS Collaboration],
  Phys.\ Rev.\ Lett.\ \ {\bf 93}, 242303  (2004).


\bibitem{KLV}H.~Kowalski, T.~Lappi, R.~Venugopalan,
  Phys.\ Rev.\ Lett.\  {\bf 100}, 022303 (2008).
  
  \bibitem{arXiv:0911.2720} 
  K.~Dusling, F.~Gelis, T.~Lappi and R.~Venugopalan,
  Nucl.\ Phys.\ A\ {\bf 836}, 159  (2010).

\bibitem{Kharzeev-Nardi}D.~Kharzeev and M.~Nardi,
  Phys.\ Lett.\ B\ {\bf 507}, 121  (2001).
  
\bibitem{arXiv:1004.3034} 
  K.~Aamodt {\it et al.} [ALICE Collaboration],
  Eur.\ Phys.\ J.\ C\ {\bf 68}, 89  (2010).

\bibitem{arXiv:1002.0621} 
  V.~Khachatryan {\it et al.} [CMS Collaboration],
  JHEP\ {\bf 1002}, 041  (2010).
  
 \bibitem{pp_mult_1}	
   K.~Aamodt {\it et al.} [ALICE Collaboration],
  Eur.\ Phys.\ J.\ C\ {\bf 68}, 345  (2010).
  
  \bibitem{pp_mult_2}
  {\bf UA5} collaboration, G.~J. Alner {\em et.~al.},  \mbox{}
  \href{http://dx.doi.org/10.1007/BF01410446}{{\em Z. Phys.} {\bf C33} (1986)
  1}; {\bf UA1} collaboration, C.~Albajar {\em et.~al.},  \mbox{}
  \href{http://dx.doi.org/10.1016/0550-3213(90)90493-W}{{\em Nucl. Phys.} {\bf
  B335} (1990) 261}; {\bf CDF} collaboration, F.~Abe {\em et.~al.},  \mbox{}
  \href{http://dx.doi.org/10.1103/PhysRevD.41.2330}{{\em Phys. Rev.} {\bf D41}
  (1990) 2330}  

\bibitem{arXiv:0808.2041} 
  B.~I.~Abelev {\it et al.} [STAR Collaboration],
  Phys.\ Rev.\ C\ {\bf 79}, 034909  (2009).
  
  \bibitem{AA_mult}
{\bf STAR} collaboration, C.~Adler {\em et.~al.},  \mbox{}
  \href{http://dx.doi.org/10.1103/PhysRevLett.87.112303}{{\em Phys. Rev. Lett.}
  {\bf 87} (2001) 112303};
{\bf PHOBOS} collaboration, B.~B. Back {\em et.~al.},  \mbox{}
  \href{http://dx.doi.org/10.1103/PhysRevLett.85.3100}{{\em Phys. Rev. Lett.}
  {\bf 85} (2000) 3100}; 
{\bf BRAHMS} collaboration, I.~G. Bearden {\em et.~al.},  \mbox{}
  \href{http://dx.doi.org/10.1016/S0370-2693(01)01333-8}{{\em Phys. Lett.} {\bf
  B523} (2001) 227}; 
 {\bf BRAHMS} collaboration, I.~G. Bearden {\em et.~al.},  \mbox{}
  \href{http://dx.doi.org/10.1103/PhysRevLett.88.202301}{{\em Phys. Rev. Lett.}
  {\bf 88} (2002) 202301}; 
 {\bf PHOBOS} collaboration, B.~B. Back {\em et.~al.},  \mbox{}
  \href{http://dx.doi.org/10.1103/PhysRevC.65.061901}{{\em Phys. Rev.} {\bf
  C65} (2002) 061901};
 {\bf PHENIX} collaboration, S.~S. Adler {\em et.~al.},  \mbox{}
  \href{http://dx.doi.org/10.1103/PhysRevC.71.034908}{{\em Phys. Rev.} {\bf
  C71} (2005) 034908}.

  
  \bibitem{arXiv:1103.1259} 
  R.~Baier, A.~H.~Mueller, D.~Schiff and D.~T.~Son,
  arXiv:1103.1259 [nucl-th].
  
  \bibitem{arXiv:1107.5296} 
  J.~-P.~Blaizot, F.~Gelis, J.~Liao, L.~McLerran and R.~Venugopalan,
  arXiv:1107.5296 [hep-ph].

\bibitem{nucl-ex/0201005} 
  B.~B.~Back {\it et al.} [PHOBOS Collaboration],
  Phys.\ Rev.\ C\ {\bf 65}, 061901  (2002).

\bibitem{arXiv:1012.1657} 
  K.~Aamodt {\it et al.} [ALICE Collaboration],
  Phys.\ Rev.\ Lett.\ \ {\bf 106}, 032301  (2011).

\bibitem{nucl-ex/0210015} 
  B.~B.~Back, M.~D.~Baker, D.~S.~Barton, R.~R.~Betts, M.~Ballintijn, A.~A.~Bickley, R.~Bindel and A.~Budzanowski {\it et al.},
  Phys.\ Rev.\ Lett.\ \ {\bf 91}, 052303  (2003).

\bibitem{arXiv:1107.4800} 
  S.~Chatrchyan {\it et al.} [CMS Collaboration],
  JHEP\ {\bf 1108}, 141  (2011).

\bibitem{Back:2003hx}
  B.~B.~Back {\it et al.}  [PHOBOS Collaboration],
  Phys.\ Rev.\ Lett.\  {\bf 93}, 082301 (2004).

\bibitem{nucl-ex/0401025} 
  I.~Arsene {\it et al.} [BRAHMS Collaboration],
  Phys.\ Rev.\ Lett.\ \ {\bf 94}, 032301  (2005).

\bibitem{d'Enterria:2010hd} 
  D.~d'Enterria, G.~K.~.Eyyubova, V.~L.~Korotkikh, I.~P.~Lokhtin, S.~V.~Petrushanko, L.~I.~Sarycheva and A.~M.~Snigirev,
  Eur.\ Phys.\ J.\ C {\bf 66}, 173 (2010).

\bibitem{arXiv:1009.0545} 
  W.~A.~Horowitz and Y.~V.~Kovchegov,
  Nucl.\ Phys.\ A\ {\bf 849}, 72  (2011).
  
  \bibitem{arXiv:1112.1061} 
  G.~A.~Chirilli, B.~-W.~Xiao and F.~Yuan,
  arXiv:1112.1061 [hep-ph].

\bibitem{arXiv:1111.2312} 
  A.~H.~Rezaeian,
  arXiv:1111.2312 [hep-ph].

\bibitem{arXiv:1007.2430} 
  E.~Levin and A.~H.~Rezaeian,
  Phys.\ Rev.\ D\ {\bf 82}, 054003  (2010).

\bibitem{arXiv:1104.3725} 
  T.~Lappi,
  Eur.\ Phys.\ J.\ C\ {\bf 71}, 1699  (2011).

\bibitem{arXiv:1011.5161} 
  J.~L.~Albacete and A.~Dumitru,
  arXiv:1011.5161 [hep-ph].

\bibitem{arXiv:1111.3031} 
  A.~Dumitru, D.~E.~Kharzeev, E.~M.~Levin and Y.~Nara,
  arXiv:1111.3031 [hep-ph].

\bibitem{arXiv:1005.0631} 
  E.~Levin and A.~H.~Rezaeian,
  Phys.\ Rev.\ D\ {\bf 82}, 014022  (2010).

\bibitem{arXiv:1001.1378} 
  J.~L.~Albacete and C.~Marquet,
  Phys.\ Lett.\ B\ {\bf 687}, 174  (2010).

\bibitem{arXiv:1110.2810} 
  J.~Jalilian-Marian and A.~H.~Rezaeian,
  arXiv:1110.2810 [hep-ph].

\bibitem{hep-ph/0506308} 
  A.~Dumitru, A.~Hayashigaki and J.~Jalilian-Marian,
  Nucl.\ Phys.\ A\ {\bf 765}, 464  (2006).

\bibitem{arXiv:1102.5327} 
  T.~Altinoluk and A.~Kovner,
  Phys.\ Rev.\ D\ {\bf 83}, 105004  (2011).
 

\bibitem{arXiv:1002.2537} 
  P.~Quiroga-Arias, J.~G.~Milhano and U.~A.~Wiedemann,
  Phys.\ Rev.\ C\ {\bf 82}, 034903  (2010)

\bibitem{arXiv:1105.3919} 
  C.~A.~Salgado, J.~Alvarez-Muniz, F.~Arleo, N.~Armesto, M.~Botje, M.~Cacciari, J.~Campbell and C.~Carli {\it et al.},
  arXiv:1105.3919 [hep-ph].

\bibitem{arXiv:1111.3646} 
  G.~G.~Barnafoldi, J.~Barrette, M.~Gyulassy, P.~Levai and V.~Topor Pop,
  arXiv:1111.3646 [nucl-th].


\end{thebibliography}
\end{document}